\documentclass[onecolumn]{emulateapj}
\usepackage{apjfonts}

\shorttitle{Quiescent X-ray Emission of Magnetars}
\shortauthors{Fern\'andez \& Thompson}	
\slugcomment{Astrophysical Journal, 660:615-640, 2007}
\begin{document}
\title{Resonant Cyclotron Scattering in Three Dimensions\\
and the Quiescent Non-thermal X-ray Emission of Magnetars}

\author{Rodrigo Fern\'andez\altaffilmark{1} and Christopher Thompson\altaffilmark{2}}
\altaffiltext{1}{Department of Astronomy and Astrophysics, University of Toronto, 60 St. George St., Toronto, ON M5S 3H8, Canada.}
\altaffiltext{2}{Canadian Institute for Theoretical Astrophysics, 60 St. George
St., Toronto, ON M5S 3H8, Canada.}

\begin{abstract}
Although the surface of a magnetar is a source of bright thermal X-rays,
its spectrum contains substantial non-thermal components.  The X-ray emission
is pulsed, with pulsed fractions that can be as high as $\sim 70\,\%$. Several
properties of magnetars indicate the presence of persistent, static currents
flowing across the stellar surface and closing within the magnetosphere. The
charges supporting these currents supply a significant optical depth to
resonant cyclotron scattering in the 1-100~keV band. Here we describe a
Monte Carlo approach to calculating the redistribution of thermal seed photons
in frequency and angle by multiple resonant scattering in the magnetosphere.
The calculation includes the full angular dependence of the cyclotron 
scattering cross section, the relativistic Doppler effect due to the motion
of the charges, and allows for an arbitrary particle velocity distribution
and magnetic field geometry.  We construct synthetic spectra and pulse profiles
for arbitrary orientations of the spin axis, magnetic axis, and line of sight,
using a self-similar, twisted dipole field geometry, and assuming that the
seed photons are supplied by single-temperature black body emission from the
stellar surface.  Pulse profiles and 1-10~keV spectra typical of AXPs are 
easily produced by this model, with pulsed fractions of $\sim 50\%$.  However,
this model cannot reproduce the hard, rising energy spectra that are observed
from SGRs during periods of activity, without overproducing the thermal
emission peak.  This suggests that the 1-100 keV emission of SGRs has a common
origin with the hard X-ray emission detected from some AXPs above 
$\sim 20$~keV.
\end{abstract}


\section{Introduction}

Magnetars have many remarkable properties that appear to derive from the decay of an
ultrastrong magnetic field.  Measurements of spindown torques and bright X-ray outbursts have revealed
dipole fields as strong as $\sim 10^{15}$ G, and imply the existence of internal fields at least
an order of magnitude stronger in some objects \citep{td96,wt06}.  More recent measurements strongly 
suggest that the external magnetic field can support electric currents that are several orders of 
magnitude larger than the Goldreich-Julian current that flows along the open field lines.  
In some magnetars, the injection of these closed-field 
currents into the magnetosphere appears to be associated with Soft Gamma Repeater burst activity; but
in others, the current appears to be sustained in the absence of bright SGR outbursts, in the class 
of magnetars known as the Anomalous X-ray Pulsars.

The basic properties of a twisted neutron star magnetosphere were outlined by \citet{tlk02}.
A detailed one-dimensional plasma model of the current in the closed magnetosphere has been 
constructed by \citet{bt06}.  The presence of such a global twist has several observational
effects, which mirror the behavior of magnetars in the following ways.

\begin{enumerate}
\item \emph{Correlation between spindown torque and X-ray hardness.} The presence of a twist
in the external magnetic field causes it to flare out slightly.  As a 
consequence, the open-field flux is larger than estimated from the dipole model,
and the spin-down torque is increased.  The charges flowing along the twisted closed field 
lines also supply a significant optical depth to resonant cyclotron
scattering.  This optical depth is independent of radius (resonant frequency) when the
magnetosphere is self-similar.  The thermal photons emitted from the surface of the star
are multiply scattered.
Thus the observed correlation 
between spin-down torque and hardness of the power-law component of the 1-10 keV X-ray spectrum 
\citep{marsden01} can be explained by adjusting a single parameter:  the angle through which the
magnetospheric field lines are twisted \citep{tlk02}.

\item \emph{Persistent changes in X-ray pulse profiles.} Following the 1998
giant flare, the pulse profile of SGR~1900$+$14 changed to a single, nearly sinusoidal
peak, from a more complicated profile with multiple peaks that was observed before
the flare \citep{woods01}. The change
in the pulse profile persisted even though the quiescent X-ray flux returned
to its pre-burst level.  Something similar happened to SGR 1806$-$20
after its 2004 giant flare, except that instead of simplifying, the pulse
profile became more complex \citep{woods06}. The AXP 1E~2259$+$586 also showed
a systematic change in pulse profile following the 2002 outburst, although
the duration of the change was shorter lived than in the SGRs \citep{woods04}.
It appears that the external magnetic field of a magnetar is reconfigured following
a period of burst activity \citep{woods01,tlk02}.

\item \emph{Persistent changes in $\dot P$}.  The spindown torque of 
SGR~1900$+$14 and SGR~1806$-$20 can show large-amplitude modulations, increasing over
the previous long term trend by at least a factor $\sim 4$ (\citealt{woods06}, and references
therein).  The increase sets in gradually several months after a period of burst activity,
and then persists for years.  Related but less dramatic effects are seen 
in the torque behavior of some AXPs \citep{gavriil04}.

\item \emph{Hard X-ray emission.}  Emission above $20$~keV has been measured 
from both SGRs and AXPs (\citealt{gotz06}, and references therein).  In the 
AXPs, the spectrum sometimes shows a strong upward break between 10 and 50 keV,
with the output at 100 keV being 10 times stronger than the thermal emission
from the neutron star surface.  It appears that the bolometric output of these
objects is dominated by the magnetosphere. The additional high-energy component
could be explained by bremsstrahlung emission from the transition layer near 
the neutron star surface, or possibly by a pair cascade at $\sim 100$ km
from the neutron star \citep{tb05}.  

\end{enumerate}

In this paper, we confront the magnetar model with some of the 
observational evidence listed above.  We focus on the 
part of the non-thermal X-ray spectrum that lies directly above
the black body peak, and also on the X-ray pulse behavior in this band 
(items 1 and 2 above).
The most robust way of testing the twisted magnetosphere hypothesis is to
produce synthetic analogs of observable quantities, using the self-similar
model developed by \citet{tlk02}. The radiation transport through the magnetosphere 
is sensitive to several geometrical effects, which include the angular emission 
pattern of cyclotron scattering by moving charges, the magnetic field geometry,
orientation with respect to the observer, and the particle velocity distribution.

A first strategy is to ignore all the geometrical complications, focusing on the 
essential physics entering the multiple scattering process in a one-dimensional geometry.
The problem of radiation transport through the accretion column of an X-ray pulsar
has been treated in this way, using both Feautrier methods to solve the radiation
transport equation \citep{meszaros85a,meszaros85b} and Monte Carlo techniques \citep{pravdo81, wang88}.
The more general problem of the transfer of X-ray photons through a single resonant surface
has been analyzed by \citet{zheleznyakov96} and \citet{lyutikov06} by restricting the
transfer of photons to one spatial dimension and assuming a sub-relativistic
and thermal distribution of scattering charges.

We do not expect narrow spectral features to form as the result of the transfer 
of thermal  X-ray photons through the quiescent, twisted magnetosphere.
The cyclotron optical depth is regulated to a value close to unity over a broad
range of frequencies.  By contrast, the optical depth is much higher in a pulsar
accretion column, and is concentrated over a narrower range of radius and magnetic field.
(In addition, the equivalent width of the absorption feature that forms in the cold atmosphere
of a passively cooling magnetar is suppressed by polarization mode exchange \citealt{holai04}.)
The optical depth varies significantly with angle in the twisted dipole magnetic field,
and the field lines impart a non-trivial angular structure to the particle velocity
distribution.  This means that a numerical simulation in three dimensions is required
to model the formation of a high-energy tail to the spectrum, and the amplitude and shape 
of the X-ray pulses that are seen by a distant observer.

In this paper, we report the development of a Monte Carlo code that follows the
transfer of X-ray photons through the stellar magnetosphere, with resonant cyclotron
scattering as the source of opacity. The code treats the scattering problem in 
three dimensions, for an arbitrary magnetic field geometry, and an arbitrary velocity
distribution of the scattering charges.  We account for the
relativistic Doppler effect and polarization exchange during scattering, but neglect
the effects of recoil (which are important only at high energies where the spectra
of AXPs and SGRs are not well measured, and where additional emission components are
sometimes observed).  The output spectra and pulse profiles are obtained for an arbitrary line of sight,
and for an arbitrary orientation of the spin axis and magnetic axis of the neutron star.

The paper is structured as follows.  In \S~2 we review the physical input to
the model, and in \S~3 we describe the Monte Carlo code.  Results are presented
and discussed in detail in \S~4.  A comparison with the existing data 
and general conclusions follow in \S~5.

\section{Resonant Cyclotron Scattering in a Non-Potential Magnetosphere}\label{s:rescyc}

In this section we describe the physical model that we are simulating.
We first review the properties of a neutron star magnetosphere that
carries a net twist \citep{tlk02}. We then review the process of
resonant cyclotron scattering and how it depends on the
structure of the magnetosphere.
At frequencies lower than the surface cyclotron frequency, the 
optical depth depends only on the twist angle and the drift speed of the
charges.   The effects of relativistic particle motion will be considered
in detail.

 In the magnetar model, there are two 
sources of keV photons.  The first is deep cooling.  As the internal magnetic field is transported outwards, part of the
field energy is converted to heat, and the remainder to a deformation of the
magnetic field outside the star.  
Part of this internal heat is conducted to the surface, and the remainder radiated as neutrinos
\citep{td96}.  This component of the blackbody flux will be carried by the E-mode, given that its
opacity is strongly suppressed in the surface layers by a factor $\sim (m_ec\omega/eB)^2
\sim (k_{\rm B}T_{\rm bb}/m_ec^2)^2\,B_{15}^{-2}$ \citep{silantev80}.

The impact of current-carrying particles on the surface is a second possible source of
kilo-electron volt photons \citep{tlk02}.  Some fraction of the particle kinetic energy
will be deposited in a layer which is optically thick
to the O-mode, but optically thin to the E-mode.  This energy would then be re-radiated in the O-mode with a nearly black body frequency distribution.
Although ions are probably able to penetrate this deep, it has been argued
that a relativistic electron (or positron) beam will deposit a signficant
fraction of its energy in an optically thin layer, which is re-emitted
in the O-mode at a much higher temperature \citep{tb05}.
A recent re-calibration of the distance to several AXPs shows that their
1-10 keV luminosities are tightly clustered around $\sim 1\times 10^{35}$ ergs s$^{-1}$ \citep{durant06b}.  These sources
have a range of spin-down rates and their X-ray spectra are of variable hardness.  The clustering of
the 1-10 keV emission therefore suggests that the rate of heat conduction to the surface is buffered by
bulk neutrino cooling, and that the black-body emission is not dominated by particle heating.  
In at least one AXP, the magnetospheric output at $\sim 100$ keV is significantly brighter than the $\sim 1$ 
keV surface emission \citep{kuiper05,kuiper06}, and is more comparable to the 
expected neutrino luminosity \citep{td96}.  This
suggests that a significant fraction of the
energy that is released by the untwisting of the internal magnetic field ends up in a quasi-static deformation
of the external magnetic field, such as we are assuming in this paper.

A photon propagating away from the surface of a magnetar can
scatter resonantly at a significant rate over some range of
its path.  If the magnetospheric particles had a narrow velocity
distribution of width $\Delta \beta c$, then the cyclotron scattering would be
localized on a `resonant surface' with a well defined radius
$R_{\rm res}$ and a thickness $\Delta R_{\rm res} \sim \Delta\beta
R_{\rm res}$.  A similar statement holds for any photon after it
has been resonantly scattered within the magnetosphere.

The transfer of thermal photons through a single warm scattering
layer (with one-dimensional electron temperature $k_{\rm B}T_\parallel \ll m_ec^2$) results in a second component
of the spectrum that is upshifted in frequency by a factor $1+2(k_{\rm B}T_\parallel/m_ec^2)^{1/2}$ 
\citep{lyutikov06}.  This analytic result is, however, restricted to one spatial dimension,
does not include bulk motion of the charges, neglects the effect of scattering at distant points
in the magnetosphere, and becomes inaccurate when the r.m.s. speed of the charges 
exceeds\footnote{Working to first order in $\beta$, one must
approximate $[1+(k_{\rm B}T_\parallel/m_ec^2)^{1/2}]/
[1-(k_{\rm B}T_\parallel/m_ec^2)^{1/2}] \simeq 1+2(k_{\rm B}T_\parallel/m_ec^2)^{1/2}$.}
$\beta \simeq 0.3$.  

In a more realistic, three-dimensional magnetosphere a 
power-law component of the spectrum results from two effects:  
\begin{enumerate}
\item Multiple resonant scattering within a single magnetic hemisphere, where the
charge flow is converging and the average increase $\Delta\omega$ in photon frequency per scattering
is first order in $\beta$; and
\item Multiple scatterings in different magnetic hemispheres, for which
$\Delta\omega \sim \beta^2$ if the particle drift speed parallel to the 
magnetic axis is symmetric in the two hemispheres.
\end{enumerate}
For example, when the charge flow in the twisted magnetosphere
is carried by electrons and positrons, each hemisphere will contain one
species of charge whose bulk motion is directed toward the star.  Even
when electrons are the only species of light charge (the positive charge
flow being carried by ions), the electron velocity will 
converge in one magnetic hemisphere.  The optical depth is independent
of frequency in a self-similar magnetosphere, and a significant
fraction of the backscattered photons undergo multiple scattering if 
the field is twisted through $\sim 1$ radian
(\citealt{tlk02}; eq. [\ref{eq:taures}]).

\subsection{Twisted Self-Similar Magnetospheres}
\label{s:twisted_field}

\citet{tlk02} 
solved the force-free equation $\mathbf{J}\times \mathbf{B} = 0$ 
outside a spherical surface with 
a self-similar ansatz \citep{lyndenbell94,wolfson95}.  These solutions
form a one-parameter sequence, labeled by the net twist
angle of the field lines that are anchored close to the magnetic poles,
\begin{equation}\label{eq:twist}
\Delta \phi_\mathrm{N-S} = 2\lim_{\theta_0\to 0}\int_{\theta_0}^{\pi/2}\frac{B_\phi(\theta)}{B_\theta(\theta)}
\frac{d\theta}{\sin{\theta}}.
\end{equation}
Here $\theta$ and $\phi$ denote the polar and azimuthal angle relative to the magnetic
axis. The magnetic field is
\begin{equation}\label{eq:bcomp}
{\bf B}(r,\theta) = {B_\mathrm{pole}\over 2}
\left(\frac{R_\mathrm{NS}}{r}\right)^{2+p} {\bf F}(\cos\theta),
\end{equation}
where $B_\mathrm{pole}$ is the field strength at the magnetic poles, 
$r$ is the radial coordinate, and $R_\mathrm{NS}$ the stellar radius.
The components of ${\bf F}$ are expressed in terms of 
a function $f(\cos{\theta})$,
which is the solution to the ordinary differential equation
\begin{equation}
\sin^2{\theta}f^{\prime\prime} + Cf^{1+2/p} + p(p+1)f =0.
\end{equation}
One finds $F_r = -f'$; $F_\theta = (p/\sin\theta)f$;
and $F_\phi = [C/p(p+1)]^{1/2}f^{1/p}\times F_\theta$.  The function
$f(\cos\theta)$ satisfies the three boundary conditions $f'(0) = 0$, $f'(1) = -2$, and $f(1) = 0$. 
These force-free equilibria\footnote{There are two solutions for each value of the constant 
$C \leq .873$, and the two branches connect smoothly at $C = .873$.  Together they form a one-parameter
sequence labeled by $p$ or $\Delta\phi_{\rm N-S}$.} 
smoothly interpolate between a dipole ($\Delta \phi_\mathrm{N-S} = 0$, $p = 1$) and a split monopole
($\Delta \phi_\mathrm{N-S}=\pi$, $p=0$). The poloidal field lines
remain close to a dipole as long as $\Delta \phi_\mathrm{N-S}\lesssim 1$.

The current density induced along the twisted field lines is \citep{tlk02}
\begin{equation}\label{eq:current}
\mathbf{J} = \sum_i Z_ie\,n_i\, \textrm{{\boldmath $\beta$}}_i
c = \frac{(p+1)c}{4\pi r}\frac{B_\phi}{B_\theta}\, \mathbf{B}
\simeq {c\Delta\phi_{\rm N-S}\over 4\pi r}\sin^2\theta {\bf B},
\end{equation}
where $Z_ie$ is the electric charge of particle species $i$, 
$n_i$ is the particle density, and 
$\textrm{{\boldmath $\beta$}}_i= \beta_i \hat B$ is the particle 
velocity directed along $\hat B = {\bf B}/B$ in units of the speed of light 
$c$.  

The closed-field current (\ref{eq:current}) implies a much higher
flux of charged particles than can be supplied by the rotationally-induced
(corotation) charge density.  As a result, the magnetospheric
plasma contains both positive and negative charges with nearly equal
densities \citep{bt06}.  One has
$J/|\rho_{\rm GJ}|c \sim \Delta\phi_{\rm N-S}\,(\Omega R_{\rm max}/c)^{-1}$,
where $\rho_{\rm GJ} = -\textrm{{\boldmath $\Omega$}}\cdot \mathbf{B}/2\pi c$
\citep{goldreich69}, {\boldmath $\Omega$} is the spin angular velocity of the star,
and $R_{\rm max} \simeq r/\sin^2\theta$ is the maximum radius of the
poloidal field line that passes through the position $(r,\theta)$ in the
magnetosphere.  We will consider both a unidirectional flow of the scattering charges 
at any given position in the magnetosphere, and also 
a bi-directional flow.  The first case corresponds to an 
electron-ion plasma,  and the second to a pair-dominated plasma.

\subsection{Basic Properties of Resonant Cyclotron Scattering of
X-ray Photons}
\label{s:resonant_scattering}

When a plasma is threaded by a magnetic field, cyclotron absorption will occur 
whenever the condition
\begin{equation}
\omega = \omega_c \equiv \frac{|Z|eB}{mc}
\end{equation}
is satisfied in the rest frame of a charged particle.  Here, $\omega$ is 
the photon frequency and $Ze$ and $m$ are the particle charge and mass.
Absorption by motionless charges occurs at a radius
\begin{equation}\label{eq:rres}
R_\mathrm{res}(\omega) = 10.5\, R_\mathrm{NS} \left[|Z|\left(\frac{B_\mathrm{pole}}{10^{14}\textrm{G}}\right)
\left(\frac{m_e}{m}\right)
\left(\frac{\textrm{1keV}}{\hbar \omega}\right) \xi\left(\theta,\Delta \phi_\mathrm{N-S}\right)
 \right]^{1/(2+p)}.
\end{equation}
For a 1 keV photon, this works out to $R_{\rm res} \sim 
200\,B_{\rm pole,15}^{1/3}$ km for electrons and $20\,B_{\rm pole,15}^{1/3}$ km
for protons.  The function $\xi$ contains the angular dependence 
of the magnetic field strength:  
\begin{equation}
\xi^2 = {1+3\cos^2\theta\over 4}
\end{equation}
for a dipole, and
\begin{equation}
\xi^2 = {1\over 4}\left[(f')^2 + \left({pf\over \sin\theta}\right)^2 + {pC\over p+1}f^{2+2/p}\right]
\end{equation}
for the self-similar twisted magnetosphere (eq. [\ref{eq:bcomp}]).  

The lifetime of the first Landau level,
$1/\Gamma = 3 m_e^3 c^5/(4e^4B^2) 
\sim 3\times 10^{-16}(m/m_e)^3(B/10^{12}\,$G)~s,
is much shorter than the dynamical time $R_\mathrm{res}/c$ 
when the cyclotron energy is greater than $\sim 10^{-3}(m/m_e)^{4/5}$~eV.
In this fast-cooling regime, the combined absorption and reemission 
can be treated as a scattering process, with a rest-frame cross section 
\begin{equation}\label{eq:sigma_res}
\sigma_\mathrm{res} = 4\pi^2 \frac{|Z|e}{B}|e_{l,r}|^2\,
\omega_c\delta(\omega-\omega_c)
\end{equation}
(e.g. \citealt{meszaros92}). 
The amplitude 
\begin{equation}
e_{l,r} = \hat e \cdot \left({\hat x \pm i\hat y\over\sqrt{2}}\right)^*
\end{equation}
is the overlap of the photon's polarization with a left ($+$) or
right ($-$) circularly polarized mode (depending on whether the scattering
charge is negative or positive).  Here $\hat x$ and $\hat y$ are unit vectors 
perpendicular to ${\bf B}$.  Integrating through the cyclotron resonance
yields a cross section that is much larger than Thomson, by a factor 
$\sim (e^2/\hbar c)^{-1} (\hbar\omega_c/m_ec^2)^{-1} \sim 10^5$ for a keV photon.
We neglect the electron recoil in our calculations, and relativistic
effects on the scattering cross section (e.g. \citealt{herold79}).  These
effects start to be significant at photon energies 
$\gtrsim 50$~keV~$\sim m_ec^2/10$ 
if the electrons are sub-relativistic.

The surface magnetic field of the AXPs and SGRs is believed to 
exceed $B_{\rm QED} = m_e^2c^3/e\hbar = 4.4\times 10^{13}$ G,
where the energy of the first electron Landau level 
begins to exceed $\sim m_ec^2$.  
We are concerned here only with resonant scattering in the parts of the
magnetosphere where the Landau levels of the scattering charges are
non-relativistic.  Resonant absorption of kilo-electron volt photons
close to the surface of a magnetar creates a high-energy gamma ray that
is converted to an $e^+-e^-$ pair near the emission site \citep{bt06}. 
The ion cyclotron energy is in the kilo-electron volt range even at
the surface of a magnetar, and so the effects of ion cyclotron scattering
are concentrated much closer to the star.
The cross sections that follow generally apply only in this
non-relativistic regime.

The current (\ref{eq:current}) supplies enough particles
for resonant cyclotron scattering to become the dominant 
source of opacity in the keV range \citep{tlk02,lyutikov06}.   
Integrating eq. (\ref{eq:sigma_res}) through the resonance gives
an optical depth
\begin{equation}\label{eq:taures}
\tau_\mathrm{res}\sim {\Delta \phi_\mathrm{N-S}\over |\beta|}\;\sin^2{\theta}.
\end{equation}
This expression does not depend on the mass or charge of the scattering 
particle, and applies as long as the resonant energy is less than
$mc^2$ and lies below the surface cyclotron energy.
This result is a direct consequence of the self-similarity of the force-free
construction.  
A more realistic magnetosphere has a more inhomogeneous distribution of twist,
contains higher-order multipoles, and is non-axisymmetric.

The resonance condition is changed by bulk motion of
the charges along the magnetic field lines, and becomes
\begin{equation}\label{eq:omega_doppler}
\omega = \omega_\mathrm{D} \equiv \frac{\omega_c}{\gamma (1-\beta \mu)}.
\end{equation}
Here $\gamma = (1-\beta^2)^{-1/2}$ is the
Lorentz factor and $\mu = \hat k\cdot\hat B \equiv \cos{\theta_\mathrm{kB}}$
is the direction cosine of the photon with respect to the magnetic field
in the stellar frame. In this case the resonant
radius in eq.~(\ref{eq:rres}) needs to be multiplied by a factor $[\gamma(1-\beta\mu)]^{-1/(2+p)}$.
Similarly, the cross section in the stellar frame equals eq.~(\ref{eq:sigma_res}) multiplied by 
$(1-\beta \mu)$ [e.g. \citealt{rybicki79}], with $\omega_c$ replaced by $\omega_\mathrm{D}$,
and with $|e_{l,r}|^2$ evaluated in the rest frame of the charge:
\begin{equation}\label{eq:sigma_resb}
\sigma_\mathrm{res} = 4\pi^2\,
(1-\beta\mu)\,
\frac{|Z|e}{B}|e_{l,r}'|^2\,
\omega_\mathrm{D}\delta(\omega-\omega_\mathrm{D})\;\;\;\;\;\;\;\;(\beta \neq 0).
\end{equation}
Here the $'$ labels the particle rest frame.

The polarization modes of a photon in a magnetized plasma depend on the relative
contributions of plasma and vacuum polarization to the dielectric tensor (e.g.  \citealt{zheleznyakov96}).   Even in the core of a cyclotron resonance,
vacuum polarization dominates in a cold plasma with a velocity
width $\Delta \beta \ll 1$ if the condition
\begin{equation}\label{eq:vacuum_pol}
\frac{\omega_P^2}{\omega^2 |\Delta \beta|}\ll \frac{1}{45\pi}\frac{e^2}{\hbar c}
\left(\frac{B}{B_\mathrm{QED}}\right)^2\;\;\;\;\;\;(B < B_{\rm QED})
\end{equation}
is satisfied.  In eq.~(\ref{eq:vacuum_pol}), 
$\omega_P = \sqrt{4\pi e^2 n_e/m_e}$ is the electron plasma frequency.
This means that 
both electromagnetic eigenmodes are then linearly polarized even at
frequencies very close to $\omega_D$.  
The extraordinary mode (E-mode) corresponds to $\hat e = \hat B \times \hat k$,
and the ordinary mode (O-mode) to $\hat e = (\hat B \times \hat k)\times \hat k$. 
This means that each polarization eigenmode will experience the same interaction 
with the gyromotion of a positron as it does with the gyromotion of an electron:
the overlap
\begin{equation}\label{circamp}\label{eq:circular_amplitude}
|e_{l,r}|^2_\mathrm{E} = 1/2;\;\;\;\;\;\;
|e_{l,r}|^2_\mathrm{O} = \cos^2{\theta_\mathrm{kB}}/2
\end{equation}
does not depend on the sense of circular polarization.  
In the case of unpolarized photons,
$|e_{l,r}|^2_\mathrm{unpol} = (1 + \cos^2{\theta_\mathrm{kB}})/4$.

Mode exchange is clearly possible during resonant scattering
(e.g. \citealt{zheleznyakov96}; \citealt{wang88}). 
The differential cross section for scattering from mode $A$ to mode $B$
is $d\sigma_\mathrm{AB}/d\Omega^\prime \propto |e_{l,r}'(\mathrm{A})|^2|e_{l,r}'(\mathrm{B})|^2$.   
The probability that any mode $A$
is converted to the extraordinary mode is
\begin{equation}\label{eq:convert}
P(A\to E) \;=\; {|e_{l,r}'(\mathrm{E})|^2\over
|e_{l,r}'(\mathrm{E})|^2 + |e_{l,r}'(\mathrm{O})|^2} \;=\; 
\frac{1}{1+\cos^2{\theta_\mathrm{kB}^\prime}},
\end{equation}
and the probability for conversion to the O-mode is
$P(A\to O) = 1-P(A\to E)$.
These branching ratios depend on
the angle (after scattering) between $\hat k$ and $\hat B$ in the charge rest frame,
\begin{equation}\label{eq:murest}
\cos\theta_{kB}' = \mu' = \frac{\mu - \beta}{1-\beta \mu}.
\end{equation}

\subsection{Basic Degrees of Freedom of the Magnetospheric Model}\label{s:degree}

The model is defined by
\begin{enumerate}
\item The twist angle $\Delta \phi_\mathrm{N-S}$ (eqs. [\ref{eq:current}],
[\ref{eq:taures}]), which labels a
one-parameter sequence of magnetic field geometries.
  We consider only axisymmetric 
field configurations in this paper.  (Our numerical implementation can in principle handle 
more complicated, non-axisymmetric, geometries.)
\item The spectral distribution of the seed photons.
We generally assume a black body distribution, defined by a single
temperature $T_{\rm bb}$.   In numerical calculations, smoother spectra below the black body
peak are obtained by inputing photons at a single frequency, and
then convolving the output spectrum with an input black body spectrum.
We show in Appendix~\ref{s:tests} that this procedure nicely reproduces the
spectrum that is obtained from drawing photons directly from a black body
distribution.
\item The polarization of the seed photons.  We calculate the output 
spectrum assuming that the seed photons are 100\% polarized
(E-mode or O-mode).
The spectrum resulting from an arbitrary admixture of
seed polarizations can be obtained by linear superposition.
\item
The angular distribution of the seed photons.
We assume that the emitting surface is homogeneous.  The
output pulse profiles and pulsed fractions depend more sensitively on
the surface temperature distribution than does the high-energy tail
of the X-ray spectrum.   We wish to define the pulse profile
shapes that can result from radiative transport in a current-carrying
magnetosphere, and the maximum pulsed fractions that can be obtained
without resort to temperature inhomogeneities across the surface.
\item
The ratio of the seed photon energy to the cyclotron energy of the
scattering charges at the magnetic pole, e.g. $k_{\rm B}T_{\rm bb}\,
(\hbar |Ze|B_{\rm pole}/mc)^{-1}$.  Although $|Ze|$, $B_{\rm pole}$,
and $m$ are all independent quantities, they enter only in this ratio.
One must, equivalently, specify the ratio of $R_\mathrm{NS}$ to $r_{\rm bb} = 
R_\mathrm{res}(\omega = k_{\rm B}T_\mathrm{bb}/\hbar)$ (eq. [\ref{eq:rres}]),
\begin{equation}\label{eq:rstar}
\frac{R_\mathrm{NS}}{r_\mathrm{bb}} = \left( 7\times 10^{-2}\right)^{3/(2+p)}\left[ 
\frac{1}{|Z|}
\left( \frac{m}{m_e} \right)
\left( \frac{k_{\rm B}T_{\rm bb}}{0.4\textrm{ keV}} \right)
\left( \frac{10^{14}\textrm{ G}}{B_\mathrm{pole}} \right)
\right]^{1/(2+p)}.
\end{equation}
Because this ratio is small for electrons and positrons, in our calculation we inject
photons moving in the radial direction from the surface of the star.  
The propagation 
of a non-radially moving photon would result in nearly radial propagation
at the point of first scattering.  
We have checked (directly by simulation) that injecting the photons
with a broad angular distributions results in practically identical spectra 
between $\sim 1$ and $\sim 50$ keV.  A broader range of injection angles
is also considered in the case of ion scattering (\S \ref{s:ionscatt}),
where it is shown that there are modest differences in the output spectrum.
\item
The momentum distribution $f(p)$ of the charge carriers.  
To explore the dependence of the results on $f(p)$,
we use three distribution functions, which are taken to be independent of
position:
\begin{enumerate}
\item[I] A mildly relativistic flow in which the particle motion is
restricted to one direction along the magnetic field 
(e.g. $\beta > 0$).  We choose a one-dimensional Boltzmann distribution 
in energy, which is labelled by a temperature $k_{\rm B}T_0 
= (\gamma_0-1)m_ec^2$,
or equivalently by a drift speed $\beta_0 = (1-\gamma_0^{-2})^{1/2}$,
\begin{equation}\label{eq:boltzone}
f(\beta\gamma) = {1\over N_\beta\, K_1(1/[\gamma_0-1])}\exp\left[-{\gamma\over {\gamma_0-1}}\right]
= {1\over N_\beta\, K_1(1/[\gamma_0-1])}\exp\left[-{\sqrt{1+(\beta\gamma)^2}\over{\gamma_0-1}}\right].
\end{equation}
Here $K_1$ is the modified Bessel function, and $N_\beta$
number of directions in which the scattering particles move along the
magnetic field.  This is a distribution in momentum $\beta\gamma$,
and is normalized in the usual manner:
\begin{equation}
\int_0^\infty d(\beta\gamma) \; f(\beta\gamma) = 1
\end{equation}
for the asymmetric distribution with $N_\beta = 1$.  
\item[II] A mildly relativistic, one-dimensional gas with the same
Boltzmann distribution as I, but now extending over positive and negative
momenta, $-\infty < \beta\gamma < \infty$.  In this case, the normalization
of $f$ is smaller by a factor of $N_\beta^{-1} = {1\over 2}$.  
This type of symmetric
distribution plausibly describes an electron-positron gas.  
In the magnetospheric circuit, the positrons are identified with the
half of the distribution function with positive velocity, and the electrons
with the particles with negative velocity.  (Both types of particle must 
exist in essentially equal numbers when the plasma is dominated by pairs, 
because it is nearly charge-neutral.)

\item[III] A broad power-law in momentum,
\begin{equation}
\label{eq:broad_dist}
f(\beta\gamma) \propto (\beta\gamma)^{\alpha},
\end{equation}
with minimum and maximum values $(\beta\gamma)_\mathrm{min}$ and $(\beta\gamma)_\mathrm{max}$, representing a
warm relativistic plasma.   
The index $\alpha = -1$ corresponds to an equal number of particles per decade of momentum. We consider only
distributions with $\alpha \leq 0$, since distributions with positive $\alpha$ are subject to 
vigorous kinetic instabilities (e.g. \citealt{kulsrud05}).
\end{enumerate}
These three families of distribution functions are characterized by the parameters 
$\left\{\beta_0\right\}$ for the Boltzmann distributions,
and  by $\left\{\beta_\mathrm{min},\gamma_\mathrm{max},\alpha\right\}$ for the
power-law distribution. In practice, the distribution functions are sampled only over a finite
range of velocities, and upper and lower cutoffs are applied 
to the admissible values of $\beta$.  
For the distributions of type I and II, these cutoffs are
taken to be very close to zero and to unity, respectively.  See \S
\ref{s:stepadapt} for further details.

\end{enumerate}

The particle distribution function 
must actually vary with position in the
magnetosphere.   
A strong drag force acts on electrons and 
positrons at $r \sim 100$ km, where their cyclotron
energy is in the keV range \citep{tb05}. As a result, 
one expects the motion of the light charges to be mildly relativistic
in this part of the magnetosphere.  Published calculations of the 
charge dynamics in the closed magnetospheric circuit \citep{bt06}
allow for pair creation but do not model the outer part 
where $B < B_{\rm QED}$.   
Close to the star, the particles
have a broad relativistic distribution, extending up to an energy
$\gamma_{\rm res} m_ec^2 \sim (\hbar eB_{\rm pole}/m_ec k_{\rm B}T_{\rm bb})$
\citep{bt06}.  At this Lorentz factor, it is possible
to spawn new electron-positron pairs through the excitation of a charge
to its first Landau state, by absorbing a photon from the peak of the
thermal radiation field.   Understanding the interplay between these
two competing effects require coupled radiation transfer and
electrodynamic calculations, which we will pursue elsewhere.

\section{Monte Carlo Simulation}\label{monte}

The random nature of photon scattering in the magnetosphere, 
and the statistical character of observable quantities 
(spectra, pulse profiles) makes the Monte Carlo technique especially 
suitable 
(see, e.g., \citealt{sobol94} for an overview of Monte Carlo methods
for Compton scattering in hot plasmas). We now present the
details of the magnetospheric model and the numerical algorithm.
The results of our calculations are reported in \S \ref{monte}.
Some validation tests are described in Appendix~\ref{s:tests}.

\subsection{Transfer of Photons through a Cyclotron Scattering Atmosphere}
\label{s:optical_depth}

We now present the basic equations that describe the transfer of radiation
through particles with an arbitrary (relativistic) distribution function.
A photon following a ray path ${\bf r} = {\bf r}_0 + \ell\hat k$
sees a differential optical depth
\begin{equation}\label{eq:taures_def}
{d\bar \tau_{\rm res}(\omega,\hat k,\mathbf{r})\over d\ell} = 
\int_{-1}^1 d\beta\,{dn_Z\over d\beta}\,
\sigma_\mathrm{res}(\omega,\hat k,\mathbf{r})
\end{equation}
for cyclotron scattering.
The point of emission of the photon is labelled
by ${\bf r}_0$ ($\ell = 0$).  The one-dimensional velocity $\beta \hat B$
of the charges has the distribution 
\begin{equation}\label{eq:nzint}
f_Z(\beta,\mathbf{r}) = n_Z^{-1}\,{dn_Z\over d\beta},
\end{equation}
where $\int_{-1}^1\, f_Z(\beta,\mathbf{r})\,d\beta = 1$.
Given that particles of charge $Ze$ carry a fraction $\varepsilon_Z$
of the current, their space density $n_\mathrm{Z}$
can be obtained from eq.~(\ref{eq:current})
\begin{equation}
\label{eq:current_sub}
\frac{|Z| e n_\mathrm{Z}}{B} = 
\varepsilon_Z\frac{(p+1)}{4\pi r |\beta|}\frac{B_\phi}{B_\theta}.
\end{equation}

We focus on a single species of scattering charge in the remainder
of this section, and drop the subscript $Z$. Radiation transfer through
an electron-positron plasma is easily described by writing 
$n = n_- + n_+$ and 
$f(\beta,\mathbf{r}) = f_-(\beta,\mathbf{r}) + f_+(\beta,\mathbf{r})$,
where the subscript labels the sign of the charge.  We will treat the
effects of ions and electrons separately in the case of an electron-ion
plasma.  In the formulae below, the factor $\varepsilon$ represents
the fraction of the current that is carried by the {\it scattering} charges:
$\varepsilon = 1$ for a pure pair plasma.
Substituting expressions (\ref{eq:sigma_resb}), (\ref{eq:nzint}), and (\ref{eq:current_sub}) into 
eq.~(\ref{eq:taures_def}) gives
\begin{equation}\label{eq:taures_averaged}
{d\bar \tau_{\rm res}\over d\ell} 
= \varepsilon{\pi(p+1)\over r}\left(\frac{B_\phi}{B_\theta}\right)\,
\left\{\int_{-1}^{1} \frac{(1-\beta \mu)
|e_{l,r}'|^2}{|\beta|}\,\omega_\mathrm{D} \delta(\omega-\omega_\mathrm{D})\,f(\beta,\mathbf{r})d\beta \right\}.
\end{equation}

At a given location in the magnetosphere, the resonance condition~(\ref{eq:omega_doppler}) can be satisfied 
only for two values of $\beta$
\begin{equation}\label{eq:beta_res}
\beta^\pm({\bf r},\hat k,\omega) = \frac{1}{(\omega_c/\omega)^2 + \mu^2}
\left[\mu \pm \left(\frac{\omega_c}{\omega}\right)
\sqrt{\left(\frac{\omega_c}{\omega}\right)^2 + \mu^2 - 1}\right],
\end{equation}
which depend on the photon frequency $\omega$ and propagation direction
$\hat k$, and the local cyclotron frequency $\omega_c$.  
We can eliminate the delta function in (\ref{eq:taures_averaged})
through the substitution
\begin{equation}
\delta(\omega-\omega_\mathrm{D}) = \frac{\delta(\beta-\beta^{\pm})}
{|\partial \omega_\mathrm{D}/\partial \beta|_{\beta^\pm}},
\end{equation}
because $\omega_\mathrm{D}$ is function of $\beta$ only for fixed $\mathbf{r}$, $\omega$, and $\hat k$.  The 
velocity-averaged optical depth is then
\begin{equation}\label{eq:taures_averaged2}
{d\bar \tau_{\rm res}\over d\ell} = \varepsilon{\pi(p+1)\omega\over r}\,
\left(\frac{B_\phi}{B_\theta}\right)\,\left[
\frac{(1-\beta^+\mu)|e_{l,r}'|^2_{\beta^+}}{|\beta^+|}\frac{f(\beta^+,\mathbf{r})}
{|\partial \omega_\mathrm{D}/\partial \beta|_{\beta^+}} + 
\frac{(1-\beta^-\mu)|e_{l,r}'|^2_{\beta^-}}{|\beta^-|}\frac{f(\beta^-,\mathbf{r})}
{|\partial \omega_\mathrm{D}/\partial \beta|_{\beta^-}} 
\right].
\end{equation}
The superscripts correspond to the two roots $\beta^\pm$.

\begin{figure*}
\vskip .4in
\plottwo{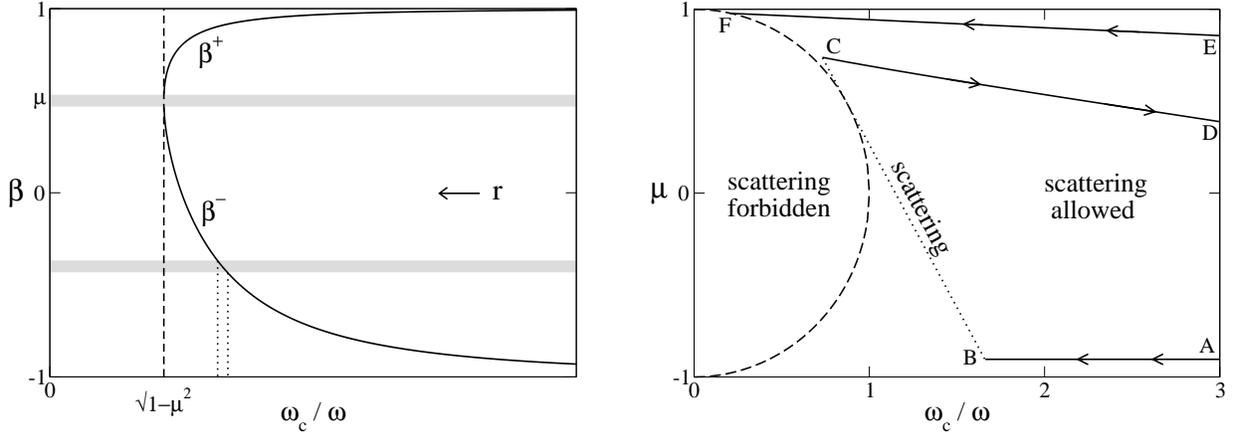}{f1b.eps}
\caption{\emph{Left panel.} Solid curve denotes the velocities
$\beta^\pm$ (eq.~[\ref{eq:beta_res}]) of the charges that are in 
resonance with a photon of frequency $\omega$ and direction cosine
$\mu$ (as measured with respect to local magnetic field).
Vertical dashed
line represents the escape surface, which is determined self-consistently
by the condition $\omega_c/\omega = \sqrt{1-\mu^2}$, or equivalently
by $\beta^\pm = \mu$ (see text).  No resonant scattering is possible to
the left of this line.   When the 
velocity distribution of the charges is narrow (horizontal shaded bars), 
resonant scattering is possible only with a narrow range of 
$\omega_c/\omega$ and of radius.
This resonant surface is even narrower in coordinate and frequency space
when it overlaps with the escape surface.
\emph{Right panel.} Sample photon trajectory in the $(\omega_c/\omega)-\mu$
plane.  Here the escape surface is represented by the dashed curve
$\mu^2+(\omega_c/\omega)^2 = 1$.
The photon trajectory starts at the surface of the star (to the extreme
right, not shown), enters the plotted domain at point A,
and scatters at point B.
After the first scattering, the frequency and direction cosine of the
photon shift over to point C.  In the example shown, the photon is backscattered
to smaller radius, and leaves and re-enters the plotted domain at points
D and E.  It finally escapes the magnetosphere at point F without any
further scatterings.}
\label{f:betares}
\end{figure*}

Once the frequency and direction of the photon are fixed, the resonance
condition can be expressed in terms of two functions $\beta^\pm$
of two local parameters $\omega_c/\omega$ and $\mu$.  A narrow
distribution of charges in velocity space implies the existence of
a well-defined `resonant surface' in coordinate space (Fig.~\ref{f:betares}).  
(The shape of this resonant surface shifts with the direction of 
the photon if the charges are in bulk motion.)
Resonant scatterings are possible only in a restricted part of this
parameter space: eq. (\ref{eq:beta_res}) has solutions only if
\begin{equation}\label{eq:escape0}
\mu^2 + \left({\omega_c\over\omega}\right)^2 > 1.
\end{equation}

The division of the photon 
trajectory into discrete steps, whose size is adapted according to 
a pre-determined velocity distribution function, is the core of the
Monte Carlo method used in this paper.
The propagation of a photon through the magnetosphere is calculated
most conveniently in this resonant velocity space:  this allows the
step size to be weighted appropriately by the density of the resonating
charges.  See the left panel of Fig.~\ref{f:betares} for an illustration.
We neglect the effect of gravity on the photon trajectory, so that
$\hat k =$ constant in between scatterings.  Applying the resonance condition 
$\omega_D = \omega$ (eq.~[\ref{eq:omega_doppler}]) gives
\begin{equation}
d\omega_\mathrm{D} = \left(\frac{\omega_\mathrm{D}}{B}\frac{dB}{d\ell} +
\frac{\partial \omega_\mathrm{D}}{\partial \mu}\frac{d\mu}{d\ell}\right)d\ell +
\frac{\partial \omega_\mathrm{D}}{\partial \beta}d\beta = 0,
\end{equation}
where $dB/d\ell = \nabla B\cdot \hat k$ and $d\mu/d\ell = 
\nabla \mu \cdot \hat k$. 
The relation between the step size in coordinate space and in velocity
space becomes
\begin{equation}\label{eq:stepsize_def}
\Delta\ell = -\frac{\left(\partial \omega_\mathrm{D}/\partial \beta \right)_{\beta^\pm}}
{\left(\partial \omega_\mathrm{D}/\partial \ell \right)_{\beta^\pm}}
\Delta\beta^\pm,
\end{equation} 
where
\begin{equation}\label{eq:dwdl}
\frac{1}{\omega_\mathrm{D}}\frac{\partial \omega_\mathrm{D}}{\partial \ell} = \frac{1}{B}\frac{dB}{d\ell} +
\frac{\beta}{(1-\beta\mu)}\frac{d\mu}{d\ell}.
\end{equation}
The differential optical depth is then 
\begin{equation}\label{eq:dtau}
\Delta \bar \tau_{\rm res} = \varepsilon{\pi(p+1)\omega\over r}
\left(\frac{B_\phi}{B_\theta}\right)\left[
\frac{(1-\beta^+\mu)|e_{l,r}'|^2_{\beta^+}}{|\beta^+|} \frac{f(\beta^+,\mathbf{r})}
{\left(\partial \omega_{D}/\partial \ell\right)}_{\beta^+}\Delta\beta^+ -
\frac{(1-\beta^-\mu)|e_{l,r}'|^2_{\beta^-}}{|\beta^-|}\frac{f(\beta^-,\mathbf{r})}
{\left(\partial \omega_\mathrm{D}/\partial\ell\right)}_{\beta^-}\Delta \beta^-\right].
\end{equation}
Since $\partial \omega_\mathrm{D}/\partial\ell$ is proportional to $\omega_\mathrm{D} =\omega$, the
optical depth is \emph{independent} of frequency. The two terms in (\ref{eq:dtau}) have opposite signs because the 
two branches of $\beta^\pm$ advance in opposite directions in 
velocity space for a given photon direction in coordinate space (see left panel of Fig.~\ref{f:betares}).
When substituting eq. (\ref{eq:stepsize_def}) into eq.~(\ref{eq:taures_averaged2}), we get a factor
\begin{equation}
\left(\frac{\partial \omega_\mathrm{D}/\partial \beta}
{|\partial \omega_\mathrm{D}/\partial \beta|}\right)_{\beta^\pm} =
\frac{\mu - \beta^\pm}{|\mu - \beta^\pm|},
\end{equation}
which is $+1$ for $\beta^-$ and $-1$ for $\beta^+$.
The gradient of the Doppler-shifted cyclotron frequency 
(eq.~[\ref{eq:dwdl}]) is generally dominated by the gradient in
magnetic field strength, and is negative for photons moving away from the star.

The photon step size $\Delta \ell$ can be chosen in such a way that
the products $f(\beta^\pm,\mathbf{r})\,\Delta\beta^\pm$ (as determined from
eq.~[\ref{eq:stepsize_def}]) yield a small value for 
$\Delta \bar\tau_{\rm res}$.  This method ensures that the bulk of the
resonant surface (where $f(\beta^\pm,\mathbf{r})$ is high)
is traversed with a small step size.  

\subsection{Photon Escape from the Magnetosphere}
\label{s:escape}

A photon cannot resonantly scatter in the outer region
where $\mu^2 + (\omega_c/\omega)^2 < 1$  (see eqs. [\ref{eq:beta_res}] and
[\ref{eq:escape0}]).  This inequality 
is a sufficient condition, which is independent of the velocity
distribution of the charges.  Nonetheless, it is possible for a photon 
to leave and then re-enter the inner region where resonant scattering 
is allowed.  To see this, it is easiest to examine the photon trajectories 
in the $(\omega_c/\omega)-\mu$
plane (the right panel of Fig.~\ref{f:betares}).  The trajectories are 
generally not straight lines in this plane, but have a mild curvature 
close to the escape surface, as long as no scattering occurs.  

We therefore adopt a second criterion, which must be satisfied if a photon
in a Monte Carlo simulation can be considered to have entirely
escaped the star and become observable.  The tangent vector of the photon
trajectory in the $(\omega_c/\omega)-\mu$ plane 
\begin{equation}\label{eq:tangent_escape}
{\bf t} = \left(\frac{\omega_c}{\omega}\right)\frac{1}{B}
\frac{dB}{d\ell}\hat \omega + \frac{d\mu}{d\ell}\hat \mu
\end{equation}
must not point in a direction that intersects the escape surface
for a second time.  [Here the unit vector $\hat \omega$ runs parallel to the
$(\omega_c/\omega)$-axis.] 
In other words, we require that ${\bf t}$ point
in the direction of the line segment $(\omega_c/\omega) = 0$,
$-1 < \mu < 1$ that forms the left boundary of the $(\omega_c/\omega)-\mu$
plane.  If this condition is not satisfied in combination with
$\mu^2 + (\omega_c/\omega)^2 < 1$, then it is necessary to follow the 
photon trajectory to check  whether it re-enters the region where
resonant scattering is allowed. 

A special case occurs for monopolar fields, or photons emitted very near
the magnetic poles in a dipole field.  In both these cases, the photon
trajectory will tend to $\mu = \pm 1$ at large distances from the star.
In numerical calculations an arbitrary cutoff needs to be taken.
In our model this is not a problem for dipole-like fields, since the
optical depth vanishes at the magnetic poles 
(recall eq.~[\ref{eq:current}]).  

The curvature of the resonant surface plays an important role in
determining the rate of photon escape.  Consider the case where
$\beta^\pm \simeq \mu$, so that the resonant surface of a photon overlaps
its escape surface.  For the purpose of illustration, imagine a top-hat
velocity distribution that is independent of position,
and has a narrow width $\Delta\beta$ centered about $\beta = \bar\beta$,
\begin{equation}\label{eq:tophat}
f(\beta,\bar \beta,\Delta \beta) = \left\{
\begin{array}{ll}
\frac{1}{\Delta \beta} & \textrm{if }(\bar \beta -\Delta \beta/2) < \beta < (\bar \beta + \Delta \beta/2);\\
0 & \textrm{otherwise}. 
\end{array}
\right.
\end{equation}
Two such distribution functions are depicted as horizontal shaded bars
in the left panel of Fig.~\ref{f:betares}. If $\bar \beta$ happens to be 
very different from $\mu$ (the lower example), then the resonant surface has a 
width 
\begin{equation}
{\Delta r\over r} \simeq \Delta\beta \left|{d\ln B\over d\ln r}\right|^{-1},
\end{equation}
which depends on $\Delta \beta$.  But if $\bar\beta - {1\over 2}\Delta\beta
< \mu < \bar\beta + {1\over 2}\Delta\beta$, then the resonant surface becomes
very thin in the radial direction, and relatively thick in the non-radial
direction.  The distance over which the photon escapes
from the resonant layer then depends on the rate at which the layer
curves away from the photon trajectory.

In practice, enough photons are scattered into an angle $\mu \simeq \bar\beta$
that the transport of photons through a resonant layer is not well
approximated by a one-dimensional model (e.g. a model in which
the cyclotron frequency has a gradient only perpendicular to the 
resonant surface: \citealt{zheleznyakov96,lyutikov06}).  
In this situation, the step size has to be
reduced appropriately (see next section).  

\subsection{Code Description: Single Photon Evolution}\label{s:code_desc}

We evolve a single photon at a time, from the point of
emission on the neutron star surface to the final escape (as defined
by the criteria \S~\ref{s:escape}).
The final frequency $\omega_{\rm out}$ and direction cosine $\mu_{\rm k}$
relative to the magnetic axis are then recorded.
This process is repeated until the distributions in 
$\omega_{\rm out}$ are $\mu_{\rm k}$ are well sampled,
according to criteria discussed in Appendix A.

In our code, a photon is a seven-dimensional vector, the
components of which are its position ${\bf r}$, direction $\hat k$,
frequency $\omega$ and polarization (E or O).\footnote{The photon's 
position and direction are 
defined in a spherical coordinate system centered on the star, 
with $\hat z$ parallel to the star's magnetic axis: 
${\bf r} = (r,~\cos{\theta}=\hat r\cdot \hat z,~\phi)$
and $\hat k = (\mu_\mathrm{k} = \hat k \cdot \hat z,~\phi_\mathrm{k})$.
Our unit of frequency is $\omega_\mathrm{bb} = k_{\rm B}T_{\rm bb}/\hbar$, 
of magnetic field $B_\mathrm{bb} = mc\omega_\mathrm{bb}/(|Z|e)$, and of length 
$r_\mathrm{bb} = R_\mathrm{res}(\omega_\mathrm{bb})$ (eq. [\ref{eq:rstar}]). }
Although the photon's electric vector is defined with respect to the
local magnetic field, the polarization will adjust adiabatically as the photon
propagates through the magnetosphere.\footnote{
The polarization tracking of an X-ray photon
breaks down only outside the last surface of cyclotron scattering.}
The code is implemented in FORTRAN77.

\begin{figure}
\vskip .4in
\epsscale{0.8}
\plotone{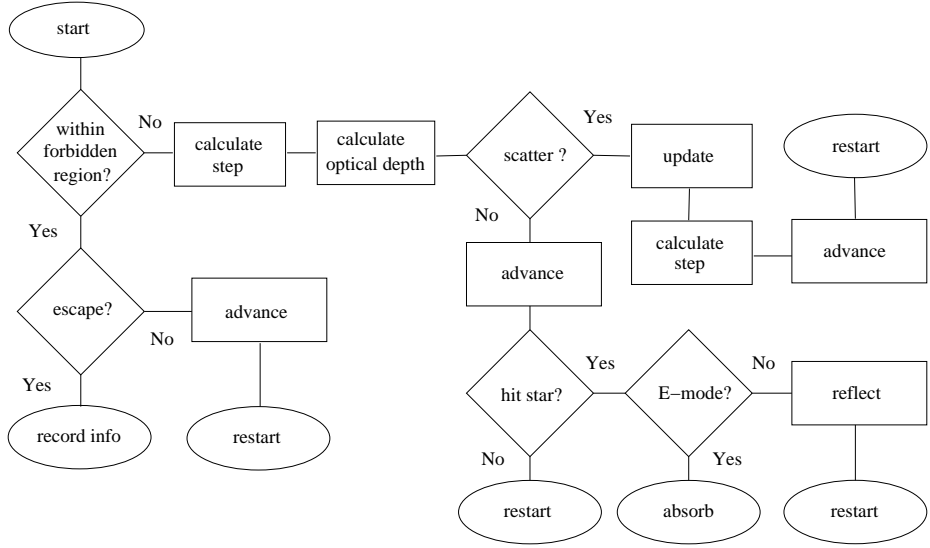}
\caption{Flow chart showing evolution of single photon. Ellipses contain starting, ending, and 
loop points in the algorithm, diamonds contain decision points, and rectangles contain simple
instructions. See text for a detailed explanation.}
\label{f:flowchart}
\end{figure}

Each photon is injected at a random point on the
stellar surface, $r=R_{\rm NS}$, with a wavevector pointing in the 
radial direction, with a frequency drawn from a black body distribution, 
and with fixed polarization.
Once the initial ray path is chosen, the photon's position and frequency are updated through a set of 
binary decisions: see the flowchart in Fig.~\ref{f:flowchart}.
Except for extreme values of the initial frequency, the scattering
rate of the photon is very small just above the point of injection,
nevertheless, we assume very small seed photon step size ($R_\mathrm{NS}/50$).

First, the subsequent step size along the photon's trajectory is determined
by an adaptive algorithm (\S \ref{s:stepadapt}) that minimizes
the step size in regions where the density of resonating charges is
highest.  The step size is 
further reduced whenever $\beta^\pm \simeq \mu$, until the condition 
\begin{equation}\label{eq:escape}
{\Delta |\mu - \beta|\over |\mu-\beta|} < 10^{-2}
\end{equation}
is satisfied.

Second, we check whether the escape criteria defined
in \S~\ref{s:escape} are satisifed.
If the photon must escape, its frequency $\omega$ and direction
angles $\mu_\mathrm{k}$ and $\phi_\mathrm{k}$ are recorded.
(The magnetic field is axisymmetric in the calculations presented in this
paper, and $\phi_{\rm k}$ is not used in the analysis.)

If the photon does not escape immediately, then we determine whether
it scatters by drawing a random number (Appendix \ref{s:stepadapt}).
The contributions of the two resonant populations of charges $\beta^\pm$ are included in the calculation
of the differential optical depth. 
If the photon scatters, then its vector is updated.
A new value for the step size is calculated, the photon is advanced, and 
the loop is restarted.

If the photon does not scatter, we advance it once step. Before restarting the loop, we check whether 
it hits the stellar surface.  An E-mode photon that hits the surface is assumed to be absorbed:
its non-resonant opacity is strongly suppressed,
and so the photon
penetrates to a depth where the absorption opacity dominates over scattering.  Such a photon does not
contribute to the output spectrum, and only its absorption is recorded.  We allow half of the O-mode photons
hitting the star to scatter elastically off the surface, and half to be absorbed.

Each photon is assumed to propagate on a straight line, and the effect of gravitational light bending 
close to the star is ignored.  This approximation is excellent near the first resonant scattering surface
if the scattering charge is an electron or positron, since $r_{\rm res} \gg 2GM_{\rm NS}/c^2 \sim 2R_\mathrm{NS}/3$ 
(see eq.~[\ref{eq:rres}]); here $M_{\rm NS}$ is the neutron star mass).   The temperature is assumed to be 
constant over the stellar surface, and so the effects of light bending can be neglected closer to the star.  
Light bending is more important when the ions are the dominant source of opacity in the magnetosphere
(as they may be in the period of strong afterglow following a Soft Gamma Repeater Bursts; \citealt{ibrahim01}).
We present some preliminary results for ion scattering in \S \ref{s:ionscatt}.
The combined effect of light bending and surface temperature variations will be included in future calculations.

This Monte Carlo procedure is independent on the magnetic field geometry and
velocity distribution function, as long as only one species of scattering particle is involved.
It can easily be generalized to include multiple scattering species, and multi-peaked velocity distribution functions (electron-positron pairs).

\subsection{Adaptive Step Size Algorithm}
\label{s:stepadapt}

The evolution of a photon's position is performed in velocity space for type I and II distributions, and in momentum space for type III (\S \ref{s:degree}).  
To maximize accuracy and minimize computing time, larger steps are
taken where $f(\beta^\pm,\mathbf{r})$ is small, and vice versa.

To implement this approach,
we assume that any velocity distribution at a given position $\mathbf{r}$
can be characterized by a single peak with half-widths $\Delta \beta_\mathrm{low}$
and $\Delta \beta_\mathrm{high}$, centered at $\beta =\bar{\beta}$. In the case of a broad distribution, 
we take the median value of $\beta$ as the central point. 
In the case where distribution function is supported over a range
of $\beta$ extending downward to 0 or upward to unity,
(e.g. the Boltzmann distribution), we restrict the values of $\beta$
at the extremes of this range to avoid numerical problems.  The limits 
are chosen empirically so that the output spectra and pulse profiles
are insensitive to the choice.   In practice, this involves sampling
some 99.8\% of the (integrated) distribution function, so that
$\int_0^{\beta_{\rm min}} f(\beta) d\beta = \int_{\beta_{\rm max}}^1 f(\beta)
d\beta = 0.001$.  
Thus, if both $\beta^+$ and $\beta^-$ are outside the resonant region,
we take a big step in 
coordinate space ($\Delta \ell = r/10$). If the value of
either $\beta^+$ or $\beta^-$ is approaching a possible resonance,  we
refine the step in velocity/momentum space until the edge of the region is resolved to a fiducial accuracy,
namely $\Delta \beta_\mathrm{low/high}/100$. Finally, if either of them lies within
the resonance surface, we set the step to a small logarithmic increment in momentum space:
\begin{equation}
d(\gamma \beta)_\mathrm{new} = \textrm{sgn}[d(\gamma \beta)_\mathrm{old}]\, |\gamma \beta|/100,
\end{equation}
or equivalently, $d\beta_\mathrm{new} = \textrm{sgn}(d\beta_\mathrm{old})\, (1-\beta^2)|\beta|/100$, where the
subscripts old and new refer to the stepsize before and after refining, and sgn is the sign function. The
numerical factor was calibrated empirically to ensure reasonable smoothness in the angular distributions
of all cases explored. For the special case $\beta^\pm = 0$, we set the step to a small constant.
Once the appropriate size of the step in either coordinate or velocity space has been 
determined, the complete set of steps ($\Delta \ell$,$\Delta \beta^+$, and $\Delta \beta^-$) is determined 
from eq.~(\ref{eq:stepsize_def}).

The differential optical depth~(\ref{eq:dtau}) is calculated by separating the contributions of the two
resonant populations $\beta^+$ and $\beta^-$ of particles, 
$\Delta \bar \tau_{\rm res} = \Delta \bar \tau^+_{\rm res} + \Delta \bar \tau^-_{\rm res}$.
Next we generate a uniformly distributed random number $\Re$ in the range $[0,1]$.
If $\Re < (1-\exp{[\Delta \bar \tau_{\rm res}]})$, the photon scatters, in
which case the new direction angles and frequency are calculated according to the method described in
\S~\ref{s:angles}. In case both $\Delta \bar \tau^+$ and $\Delta \bar \tau^-_{\rm res}$ are non-zero, the value of $\beta$
used for scattering is chosen by generating a new random number $\Re_1$; if
$\Re_1 < \Delta \bar \tau^+_{\rm res}/\Delta \bar \tau_{\rm res}$, the photon scatters with $\beta^+$, 
otherwise it scatters with $\beta^-$.

\subsection{Choice of New Photon Direction after Scattering}
\label{s:angles}

The direction cosine of the photon is chosen randomly
in the rest frame\footnote{Througout this subsection, the prime labels
the photon after scattering.} of the scattering particle, with a weighting
appropriate to the particular scattering process.  We use a similar
approach to calculate the surface reflection of an O-mode photon.

\begin{figure}
\vskip .4in
\label{f:scattering_reflection}
\epsscale{0.4}
\plotone{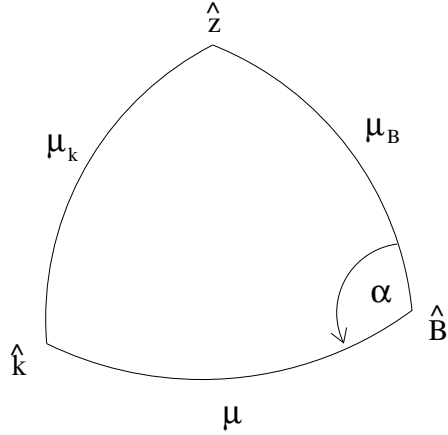}
\caption{ 
Angles defined at a scattering site, in the stellar frame. The 
vectors $\hat z$, $\hat B$, and $\hat k$ correspond to the directions of
the magnetic axis, the local magnetic field, and the photon wavevector, 
respectively. See text for the definition of 
other quantities. 
}
\end{figure}

The differential cross section for resonant cyclotron scattering 
into {\it both} polarization modes is proportional to
\begin{equation}
\label{eq:dsigmadomega_scattering}
\frac{d\sigma_{\rm res}}{d\mu'd\alpha}\propto 1 + \mu'^2,
\end{equation}
(e.g. \citealt{wang88}).  Here $\mu' = \hat B \cdot \hat k$ 
is the direction cosine of the outgoing photon in the rest frame
of the scattering particle, and $\alpha$ is its azimuthal angle
about the local magnetic field direction. 
The cumulative probability of scattering 
into an angle less than or equal to $\mu'$ is then
\begin{equation}
P(\mu') = \frac{1}{8}\left(\mu'^3 + 3\mu' + 4\right). 
\end{equation}
We set $P$ equal to a uniformly distributed random number in the range $[0,1]$,
and then solve for the value of $\mu'$.  The direction cosine 
$\mu$ of the photon in the rest frame of the star is obtained
from eq. (\ref{eq:murest}), since the resonant value of $\beta$ is known.

The polarization of the outgoing photon is then determined by 
drawing a second random number in the range $[0,1]$:  the
outgoing photon is in the E-mode if this number is smaller than 
$(1+\mu'^2)^{-1}$, and is in the O-mode otherwise.

The azimuthal angle $\alpha$ is uniformly distributed between 0 and $2\pi$;
we take the zero of $\alpha$ to coincide with the $\hat B - \hat z$ plane
(see Fig.~\ref{f:scattering_reflection}). 
The angle $\mu_\mathrm{B} = \hat B \cdot \hat z$ is known from the position 
and magnetic field geometry.
We then find the magnetic polar angle $\mu_\mathrm{k}$ of the outgoing 
photon using the cosine theorem for spherical triangles,
\begin{equation}
\mu_\mathrm{k} = \mu_\mathrm{B}\mu + \sqrt{(1-\mu_\mathrm{B}^2)(1-\mu^2)}\cos{\alpha}.
\end{equation}
To find the azimuthal angle $\phi_\mathrm{k}$ about the magnetic axis
of the star, we solve for the cartesian components
$k_\mathrm{x}$, $k_\mathrm{y}$, and $k_\mathrm{z}$, using the system of equations
\begin{eqnarray}
k_\mathrm{z} & = & \mu_\mathrm{k}\\
\hat k \cdot \hat B & = & \mu\\
\label{eq:kdotBcrossz}
\hat k \cdot (\hat B \times \hat z) & = & \sqrt{(1-\mu_\mathrm{B}^2)(1-\mu^2)}\sin{\alpha}.
\end{eqnarray}
This gives $\phi_\mathrm{k} = \arctan{(k_\mathrm{y}/k_\mathrm{x})}$.

\subsection{Statistical Treatment of Recorded Data}
\label{s:stat_treatment}

The state of each escaping photon is defined by its frequency $\omega$ 
and direction cosine $\mu_\mathrm{k}$ relative to the magnetic axis.
(The magnetosphere and photon source are axially symmetric, and so
we ignore the azimuthal angle $\phi_\mathrm{k}$.)

Care must be taken in choosing the frequency distribution of the seed
photons.
It is straightforward to draw directly from a blackbody number distribution
\begin{equation}
\label{eq:bb_number}
B\left(\log{\left[\frac{\omega_\mathrm{in}}{\omega_\mathrm{bb}}\right]}\right) =
\frac{dN}{d\log{(\omega_\mathrm{in}/\omega_\mathrm{bb})}} \propto
\frac{(\omega_\mathrm{in}/\omega_\mathrm{bb})^3}{\exp{(\omega_\mathrm{in}/\omega_\mathrm{bb})}-1},
\end{equation}
where $\omega_\mathrm{in}/\omega_\mathrm{bb}$ is the input frequency in units of 
$\omega_\mathrm{bb} \equiv k_{\rm B}T_{\rm bb}/\hbar$. However, it is difficult to obtain good
sampling of this distribution at very low and very high frequencies, because it
drops off rapidly at $\omega \ll \omega_{\rm bb}$ and $\omega \gg \omega_{\rm bb}$.

We therefore adopt an alternative procedure: we use the same input frequency for all the
photons, and store the output frequencies and direction angles in a normalized \emph{response function}
\begin{equation}
\label{eq:response}
R\left(\log{\left[\frac{\omega_\mathrm{out}}{\omega_\mathrm{in}}\right]},\mu_\mathrm{k} \right) =
\frac{1}{N_\mathrm{phot}}\frac{dN}{d\mu_\mathrm{k} d\log{(\omega_\mathrm{out}/\omega_\mathrm{in})}}.
\end{equation}
Here $\omega_\mathrm{out}/\omega_\mathrm{in}$ is the frequency shift relative to the input
frequency, and $N_\mathrm{phot}$ is the total number of photons. The outgoing histogram
in frequency and direction is then
\begin{equation}
\label{eq:histogram}
H\left(\log{\left[\frac{\omega_\mathrm{out}}{\omega_\mathrm{bb}}\right]},\mu_\mathrm{k} \right) =
\int R\left(\log{\left[\frac{\omega_\mathrm{out}}{\omega_\mathrm{in}}\right]},\mu_\mathrm{k} \right)\,
B\left(\log{\left[\frac{\omega_\mathrm{in}}{\omega_\mathrm{bb}}\right]}\right)
\, d\log{\left[\frac{\omega_\mathrm{in}}{\omega_\mathrm{bb}} \right]},
\end{equation}
which is a convolution in $\log{\omega}$. The underlying assumption in this procedure
is that the outgoing distribution in $\mu_\mathrm{k}$ is independent of frequency, which certainly
holds for the self-similar field geometry, but not for more complicated field
geometries with multipolar structure. The drawback of this approach is that by choosing
a single input frequency, one implicitly fixes the the initial scattering radius
(which depends on frequency) relative to the stellar radius.  
Nevertheless, since
most photons are emitted at frequencies near the black body peak, excellent accuracy is
obtained by choosing the peak frequency 
\begin{equation}
\label{eq:omega_peak}
\omega_{\rm in} = \omega_\mathrm{peak}\equiv 2.82144 \left({k_{\rm B}T_{\rm bb}\over\hbar}\right)
\end{equation}
as the input frequency.
In Appendix~\ref{s:tests}, we describe validation runs that confirm that this
procedure produces results very
close to those obtained by drawing frequencies directly from 
the black body distribution.

The response function $R$ is normalized to unity (modulo the small effect
of photon re-absorption at the surface), because the total number of photons 
is conserved by scattering.   The upscattering of photons 
in frequency therefore implies a transfer of kinetic 
energy from the charges to the photons, which ultimately is drawn
from the energy stored in the external toroidal field \citep{tlk02, bt06}.

We now show how pulse profiles and line-of-sight dependent spectra may be
obtained from the function $H$.  The energy flux at a fixed magnetic polar
angle and in a fixed band $\omega_{\rm min} < \omega < 
\omega_{\rm max}$ is obtained by integrating $H$ over frequency,
\begin{equation}
\label{eq:dFdmuk}
\frac{dF}{d\mu_\mathrm{k}} = \int_{\omega_{\rm min}}^{\omega_{\rm max}}
\hbar \omega\, H\left(\omega,\mu_\mathrm{k} \right) \, d\log{\omega}.
\end{equation}
To simulate a pulse profile, we pick the angle $\theta_{\Omega}$ between
the rotation axis and the magnetic axis, and the angle $\theta_{\rm los}$
between the rotation axis and the light of sight.  In a coordinate system
that is frozen into the star and aligned with its magnetic axis, the rotation axis
rotates in azimuth,
\begin{equation}
\phi_\Omega = -\Omega t.
\end{equation}
To obtain an accurate pulse profile with minimal noise, we integrate
the energy flux over a beam of a finite angular width 
$\theta_\mathrm{beam} = \cos^{-1}\mu_{\rm beam}$ that is centered 
on the direction $\hat O$ to the fiducial observer,
\begin{equation}
\label{eq:pulse_profile}
F_\Omega
= \frac{1}{2\pi}\int_0^{2\pi}\int_{-1}^{1} \frac{dF}{d\mu_\mathrm{k}}\,
\Theta(\hat r^\prime \cdot \hat O - \mu_\mathrm{beam})\, 
d\mu^\prime d\phi^\prime.
\end{equation}
Here, $\Theta$ is the step function.  The band energy flux so
constructed depends only on the rotational phase $\phi_\Omega$
and the orientation angles $\theta_\Omega$, $\theta_{\rm los}$.
The angular integral (\ref{eq:pulse_profile}) is performed most
easily in an inertial coordinate system that is aligned with the
rotation axis, and in which  $\hat O = (\theta_{\rm los},0)$. 
Each point $\hat r' = (\cos^{-1}\mu',\phi')$ in the integral 
must be related to the magnetic coordinate system,
\begin{equation}
\mu_\mathrm{k} = \mu_\mathrm{\Omega}\mu^\prime + \sqrt{(1-\mu_\mathrm{\Omega}^2)
(1-\mu^{\prime 2})}\cos{(\phi_\mathrm{\Omega}-\phi^\prime)},
\end{equation}
and to the observer's direction,
\begin{equation}
\hat r^\prime \cdot \hat O 
= \mu_\mathrm{los} \mu^\prime + \sqrt{(1-\mu_\mathrm{los}^2)
(1-\mu^{\prime 2})}\cos{\phi^\prime}.
\end{equation}
In these expressions, 
\begin{equation}\label{eq:mudef}
\mu_\mathrm{los} = \cos(\theta_{\rm los}); \;\;\;\;\;\;
\mu_\mathrm{\Omega} = \cos(\theta_{\Omega}).
\end{equation}

The rotationally-averaged spectrum is obtained by removing the
frequency integral and then integrating over rotational phase,
\begin{equation}
\omega F_\mathrm{\omega} = \frac{1}{(4\pi)^2}\int_0^{2\pi}\int_0^{2\pi}\int_{-1}^{1}\,
\hbar \omega\,
H\left(\omega,\mu_\mathrm{k} \right)\,
\Theta(\hat r^\prime \cdot \hat O - \mu_\mathrm{beam})\, d\mu^\prime d\phi^\prime d\phi_\mathrm{\Omega}. 
\end{equation}

\section{Results}

Our main goal in this paper is to determine the basic types of non-thermal
spectra and pulse profiles that result from the transfer of X-ray photons
through the scattering corona of a magnetar.  To this end, we consider
only the simplest distribution of seed photons - a single temperature
black body - and the simplest example of a twisted magnetosphere,
the self-similar sequence (\ref{eq:bcomp}) with a magnetic pitch angle 
that is independent
of radius.  We only make qualitative comparisons of our calculations
to the observed population of SGRs and AXPs.
For example, features in individual pulse profiles which are not reproduced 
by our calculations could plausibly indicate the presence of hotspots on the 
stellar surface, or some non-axisymmetry in the current flowing out to
$\sim 100$ km from the star.  We are able to reproduce the 1-10 keV
pulsed fractions that are seen in all but one source 
(1E 1048.1$-$5957; \citealt{tiengo05}).

We do explore in some detail how the X-ray spectra and pulse profiles
depend on the momentum distribution $f(p)$ of the magnetospheric
particles.  We consider the two basic types of distribution function
itemized in \S \ref{s:degree}: the relativistic Boltzmann distribution
(eq.~[\ref{eq:boltzone}]); and a broad,
power-law distribution in momentum (eq.~[\ref{eq:broad_dist}]).
We also perform some test calculations (Appendix \ref{s:tests})
with a top-hat velocity distribution (\ref{eq:tophat}).
We explore a range of mean drift speeds and power-law indices, and also
vary the magnetospheric twist angle $\Delta \phi_{\rm N-S}$ 
(eq. [\ref{eq:twist}]) and the orientation angles $\mu_\Omega$ and 
$\mu_\mathrm{los}$ (eq. [\ref{eq:mudef}]).  The scattering charges are
taken to have the charge/mass ratio of the electron, with the exception
of the simulations of ion scattering presented in \S~\ref{s:ionscatt}.
The results depend weakly on whether the photons are injected in the E-mode
or O-mode (\S~\ref{s:polarization}).  

\subsection{A Simple Model of Black Body $+$ High-Energy Power-law}

The X-ray spectra of the AXPs can be fitted by a superposition of a blackbody
and a power-law, with a photon index in the range\footnote{The photon 
index $\Gamma$ is defined by $dN/d\omega \propto \omega^{-\Gamma}$, where 
$dN/d\omega$ is the number of photons emitted per unit frequency.}
$\Gamma = 2-4$ \citep{wt06}.  Although a single power-law typically
provides a good fit to persistent SGR spectra, the Galactic SGR sources 
are more distant than the AXPs and their spectra are more heavily absorbed.
A blackbody component is present during transient X-ray afterglows that are 
observed following bursts of intermediate energy (e.g. \citealt{ibrahim01}),
and in the persistent X-ray emission of SGR 1900$+$14 \citep{esposito06}.

In AXP spectra, the thermal and non-thermal components usually have a similar
normalization at an energy of 1 keV.  In other words, the power-law component
pivots about the black body peak, a behavior that was observed directly
during a $\sim 1$-day X-ray outburst of 1E 2259$+$586 \citep{woods04}.
Within that outburst, which produced more than 100 SGR-like bursts of
a sub-second duration, the high energy photon index was observed to decay
from $\Gamma = 2$ to $\Gamma = 3.5$.

This behavior suggests that the high-energy extension of the black body peak
in the X-ray 
spectrum is created by upscattering the thermal photons in a corona above the 
surface of the neutron star.  The model examined in this paper has this
property.  It also has the advantage that no fine tuning is required for
the optical depth of the scattering charges, or their mean energy.  The
optical depth is close to unity if the magnetosphere (or, more precisely,
the part of the magnetosphere 100 km distant from the magnetar surface)
is twisted.  

The standard blackbody $+$ power law fit to magnetar X-ray spectra fails
to account for an important feature of our model:  {\it the high-energy power
law component of the spectrum is not expected to continue below the
black body peak}.  This is not just an academic issue:  in sources
with relatively low extinctions, the power-law component dominates
below the thermal peak, and significantly biases the fit to the black body
temperature (e.g. the AXP XTE J1810$-$197;  \citealt{halpern05}).

This leads us to propose a simple generalization of
the standard blackbody $+$ power law model.
This model assumes that a fraction $f_\mathrm{unscatt}$ of the 
seed photons leaves the star without scattering, and the rest are 
upscattered following a power-law distribution (Appendix A).  
The angle averaged response function, eqn.~(\ref{eq:response}), can
be written as
\begin{equation}
R\left(\log{\left[\frac{\omega_\mathrm{out}}{\omega_\mathrm{in}}\right]}\right) = 
f_\mathrm{unscatt}\, \delta\left( \log{ \left[ \frac{\omega_\mathrm{out}}{\omega_\mathrm{in}}\right]}\right) +
f_\mathrm{pl} \Theta\left(\frac{\omega_\mathrm{out}}{\omega_\mathrm{in}}-1\right)
\left[\frac{\omega_\mathrm{out}}{\omega_\mathrm{in}}\right]^{\Gamma-1},
\end{equation}
where $\Theta$ is the step function. Using eqns.~(\ref{eq:bb_number}) and (\ref{eq:histogram}), we get
\begin{equation}
\label{eq:fit}
F_\omega^\mathrm{fit} 
= f_\mathrm{unscatt}\, B\left( \frac{\omega}{\omega_\mathrm{bb}}\right)
+ f_\mathrm{pl}\int_0^\omega B \left( \frac{\omega\,'}{\omega_\mathrm{bb}}\right) 
\left[\frac{\omega\,'}{\omega}\right]^{\Gamma-1} \frac{d\omega\,'}{\omega\,'},
\end{equation}
where $B(\omega/\omega_\mathrm{bb})$ is the Planck function, eqn.~(\ref{eq:bb_number}).
Even though this behavior of the response function is seen only in simulations with a monoenergetic distribution
function, eqn.~(\ref{eq:tophat}), this generic fit worked relatively well for most of our results\footnote{
This implies that $f_\mathrm{unscatt}$ is a fit parameter which may differ from the actual fraction of
unscattered photons.}.

\subsection{Photon Index}\label{s:pindex}
\label{s:fit}

\begin{figure}
\vskip .4in
\plotone{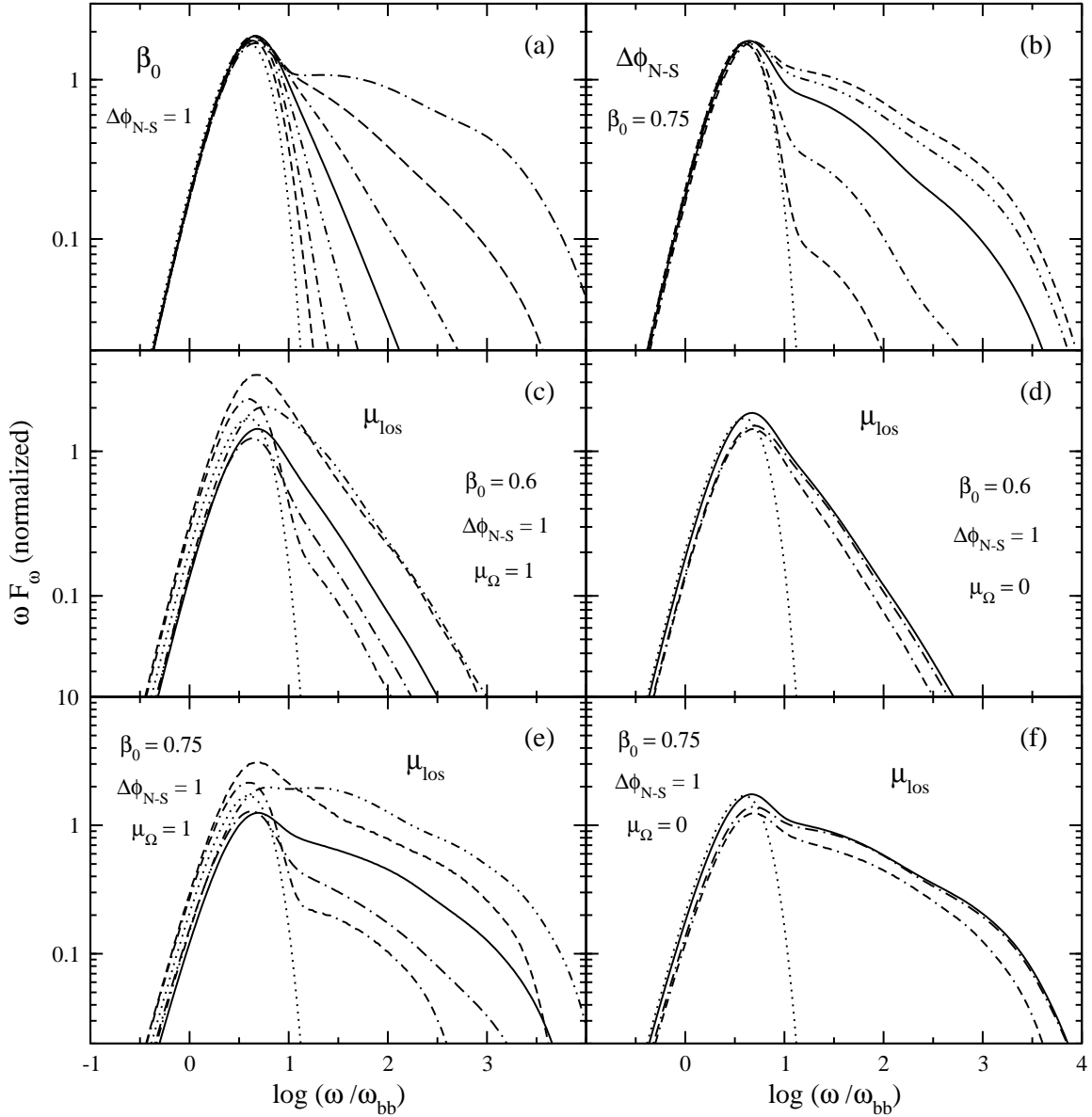}
\caption{Energy spectra obtained with a unidirectional Boltzmann
particle distribution (type I; eq.~[\ref{eq:boltzone}]). 
The unit of frequency is $\omega_\mathrm{bb} = k_\mathrm{B}T_{\rm bb}/\hbar$,
where the black body temperature $T_{\rm bb}$ is left arbitrary.
The dotted line denotes a Planck function of temperature $T_{\rm bb}$, with a
unit normalization below the curve.  Curves in panel (a) correspond
to different velocity spreads
$\beta_0 = \{0.2$ (short dash), $0.3$ (dot-short dash), $0.4$ (dot-dot-dash), $0.5$ (solid), 
$0.6$ (dash-dash-dot), $0.7$ (long dash), $0.8$ (dot-long dash) $\}$.  All curves assume that
$\Delta \phi_\mathrm{N-S} = 1$ and that the magnetic axis and line of sight 
are both orthogonal to the rotation axis ($\mu_\Omega = \mu_\mathrm{los} = 0$).  
Curves in panel (b) correspond to different twist angles,
$\Delta \phi_\mathrm{N-S} = \{0.1$ (short dash), $0.3$ (dot-dash),
$0.7$ (solid), $1$ (dot-dot-dash), $1.3$ (dash-dash-dot) $\}$.  Velocity spread $\beta_0 = 0.75$ is 
assumed and other parameters are as in (a).   Changing the line of sight direction results in
spectra shown in panels (c)-(f):  curves correspond to $\mu_\mathrm{los} = \{1$ (dash-dash-dot), 
$1/\sqrt{2}$ (dot-dash), $0$ (solid), $-1/\sqrt{2}$ (dot-dot-dash), $-1$ (dash) $\}$.
Panels (c) and (e) correspond to an aligned rotator ($\mu_\Omega = 1$), and panels (d) and
(f) to an orthogonal rotator.  Panels (c) and (d) assume a velocity spread
$\beta_0 = 0.6$, and (e) and (f) assume $\beta_0 = 0.75$.  Other parameters are the same
as in panel (a).
Photon indices obtained by fitting a blackbody plus power law function in the
frequency range $\log{(\omega/\omega_\mathrm{bb})}=0.7-1.4$ are listed in Table~\ref{t:pars_max}.
}
\label{f:wFw_sim}
\end{figure}

\begin{figure}
\vskip .4in
\plotone{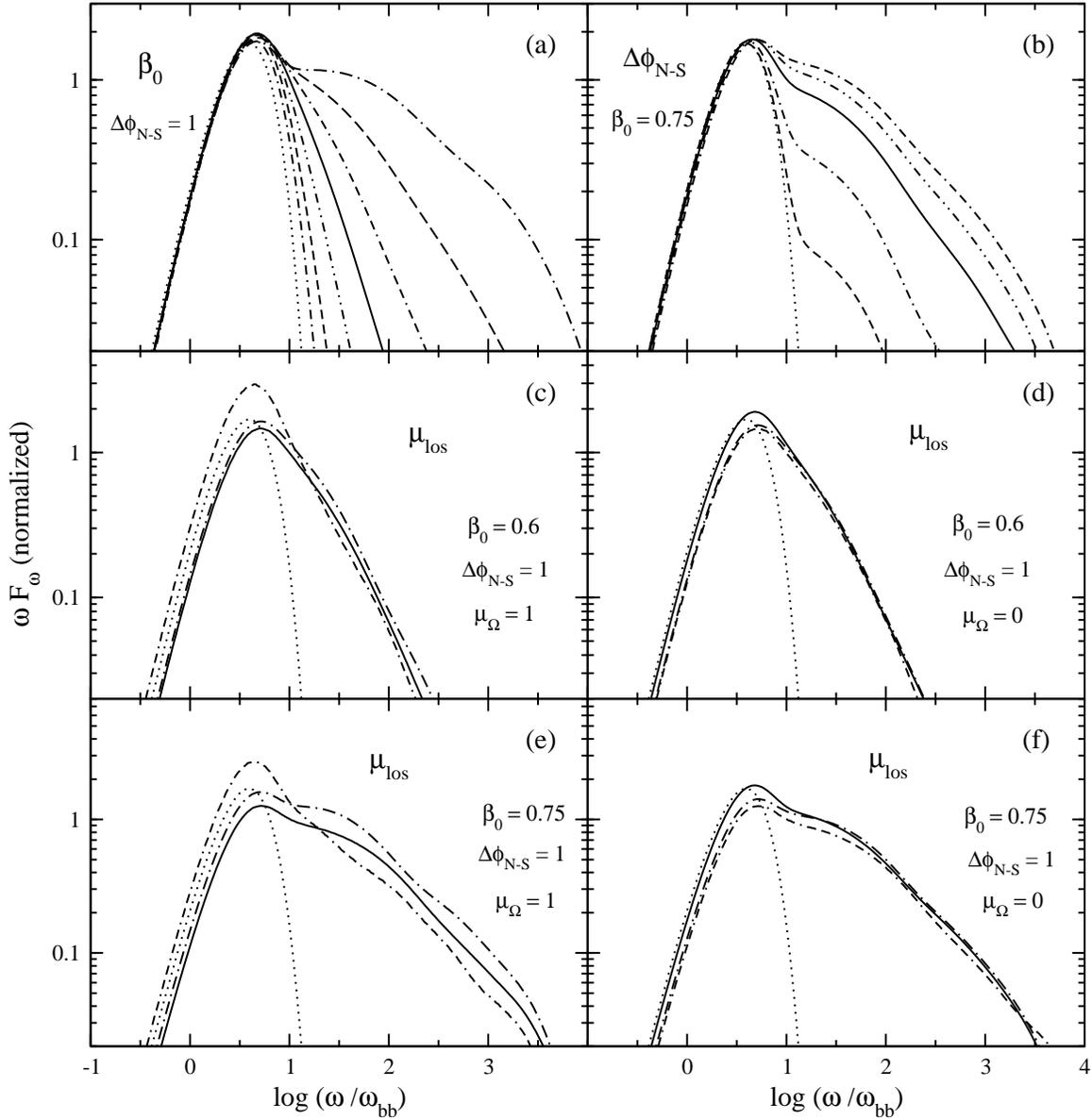}
\caption{Same as Fig.~\ref{f:wFw_sim}, but for a 
distribution function that is bi-directional,
corresponding to a pair plasma (type II; eq.~[\ref{eq:boltzone}]). 
In this case, there is a symmetry between positive and negative 
orientation angles $\mu_{\rm los}$, and curves are plotted only
for $\mu_{\rm los} > 0$.  
Photon indices are listed in Table~\ref{t:pars_bimax}.}
\label{f:wFw_e+e-}
\end{figure}

\begin{figure}
\vskip .4in
\plotone{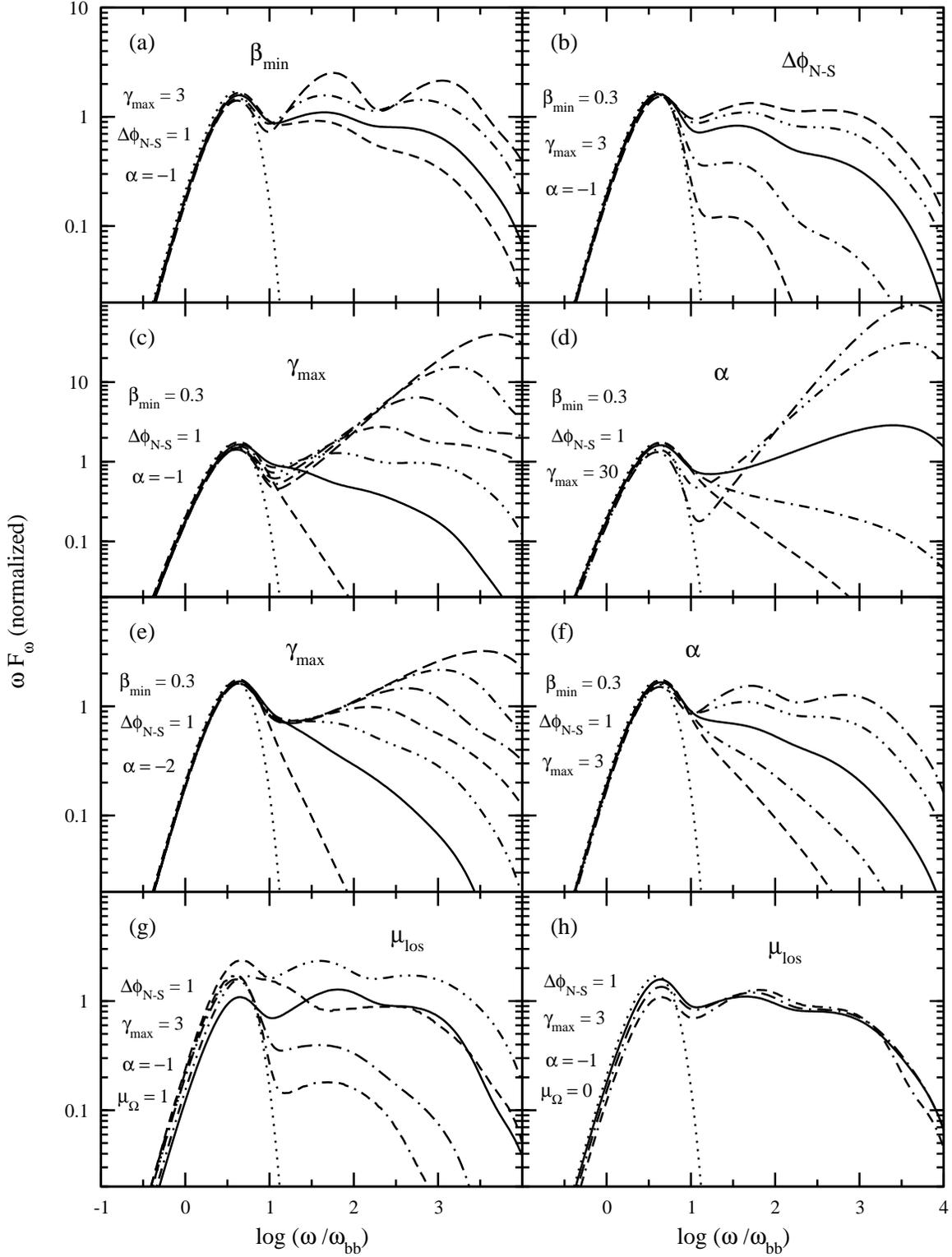}
\caption{Energy spectra obtained with the broad momentum distribution, eq.~(\ref{eq:broad_dist}).
Normalization and frequency units are the same as Fig.~\ref{f:wFw_sim}. 
Panel (a) shows results obtained with different $\beta_\mathrm{min}$, for $\gamma_\mathrm{max} = 3$, $\alpha = -1$,
$\Delta \phi_\mathrm{N-S} = -1$, and an orthogonal configuration, $\mu_\Omega = \mu_\mathrm{los} = 0$.
Curves correspond to $\beta_\mathrm{min} = \{ 0.1$ (short dash), $0.3$ (solid), $0.7$ (dot-dash),
$0.9$ (long dash)~$\}$. 
Panel (b) shows the result of varying the twist angle $\Delta \phi_\mathrm{N-S}$, for 
$\beta_\mathrm{min}=0.3$, $\gamma_\mathrm{max} = 30$, $\alpha =-1$, and the same orthogonal
configuration as (a). Curves correspond to $\Delta \phi_\mathrm{N-S} = \{0.1$ (short dash),
$0.3$ (dot-dash), $0.7$ (solid), $1$ (dot-dot-dash), $1.3$ (long dash) $\}$.
Panel (c) shows the result of varying $\gamma_\mathrm{max}$, for
$\beta_\mathrm{min} = 0.3$ and identical parameters as (a). Curves correspond to
$\gamma_\mathrm{max} = \{1.1$ (short dash), $1.96$ (solid), $3.48$ (dot-dot-dash),
$6.19$ (dash-dash-dotted), $11$ (dot-long dash), $19.6$ (dot-short dash), $34.8$ (long dash)~$\}$.  
This calculation is repeated in panel (e) for a softer particle distribution with index $\alpha = -2$.
Panel (d) shows results obtained by varying the exponent in the distribution function $\alpha$,
for $\gamma_{\rm max} = 30$ and other parameters the same as in (c). Curves correspond to 
$\alpha = \{-4$ (short dash), $-3$ (dot-short dash), $-2$ (solid), $-1$ (dot-dot-dash), 
$0$ (dot-long dash)$\}$.  This calculation is repeated in panel (f) for $\gamma_{\rm max} = 3$.
Finally, panels (g) and (h) show the aligned and orthogonal
rotator configurations, respectively, analogous to those in Fig.~\ref{f:wFw_sim} (c)-(f), but 
with simulation parameters $\beta_\mathrm{min} = 0.3$, $\gamma_\mathrm{max} = 30$, $\alpha = -1$, 
and $\Delta \phi_\mathrm{N-S} = 1$. 
}
\label{f:wFw_broad}
\end{figure}

Figures~\ref{f:wFw_sim}-\ref{f:wFw_broad} show the energy spectra ($\omega F_\omega$) 
that result from choosing particle distribution functions of types I-III, respectively. The Planck function is plotted
for comparison, and is normalized to have a unit area.  A high-energy excess above
the black body is present, and in most cases is seen to have an approximately power-law shape.  

For each output spectrum, the model described in \S \ref{s:fit} was fit
to the data using a Levenberg-Marquardt fitting routine \citep{press92} 
in the frequency range $-2 < \log{(\omega/\omega_\mathrm{bb}^\mathrm{sim})} 
< 1.7$.  This fit involved the photon index $\Gamma$, 
a thermal frequency $\omega_{\rm bb}^{\rm fit}$ (which can differ
from the input thermal frequency $\omega_{\rm bb}^{\rm sim} = k_{\rm B}T_{\rm bb}/\hbar$ by up to $\sim 10\,\%$), and the 
normalizations of the thermal (unscattered) and non-thermal (scattered)
terms in eq. (\ref{eq:fit}).  These parameters are collected in 
Tables~\ref{t:pars_max}-\ref{t:pars_broad}  (we only tabulate the
relative normalization $f_\mathrm{pl}/f_\mathrm{unscatt}$).  
In some cases, a better fit was obtained by setting $f_\mathrm{unscatt} = 0$ 
in eqn.~(\ref{eq:fit}), in which case $f_\mathrm{pl}$ is an overall 
normalization. In general, most of the fits deviate by about 10\% for 
$\log{(\omega/\omega_\mathrm{bb}^{\rm sim})} \lesssim -1$, are accurate
to a few percent in the range $-1 \lesssim \log{(\omega/\omega_\mathrm{bb}^{\rm sim})} \lesssim 1$, and 
deviate by a varying amount (1-100\%) at higher frequencies. 
The label ``poor fit" was added in Tables 1 and 3 when deviations from the 
model in the higher-frequency range exceeded $\sim 50\%$.  This is usually the
case when the resonant optical depth is small.

The hardness of the high-energy spectrum correlates with twist angle, to
which the optical depth is directly proportional, and with drift velocity of the charge carriers.
The dependence of the optical depth on the latter parameter is non-linear, often coupled with
geometric effects, which results in a stronger dependence of the photon index on the velocity
distribution parameters than on the twist angle. 
A mildly relativistic Boltzmann particle distribution function (types I-II)
yields photon indices from $\Gamma = 2$ (for $\beta_0 = 0.8$) to $\Gamma = \gtrsim 7$ 
(for $\beta_0 = 0.2$ and asymmetric/symmetric particle flows along ${\bf B}$).  
Reducing the twist from  $\Delta\phi_{\rm N-S} = 1$ results in softer 
spectra.  The choice of a broad relativistic distribution (III) produces
approximately flat energy spectra above the black body peak for 
particle index $\alpha = -1$ and 
$\gamma_{\rm max} = 3$, and for $\alpha = -2$ and a broader range of
$\gamma_{\rm max}$. Even harder spectra can be obtained by increasing the mean drift 
speed of the charges, but significant deviations from a power law appear in the high-energy tail.  

\begin{deluxetable}{cccclccccccc}
\tablecaption{Spectral Fit Parameters and Pulsed Fractions for Type I
Velocity Distribution (Boltzmann, unidirectional), 
$\Delta\beta = 0.1$\label{t:pars_max}}
\tablewidth{0pt}
\tablehead{
\colhead{$\beta_0$} & 
\colhead{$\Delta \phi_\mathrm{N-S}$} & 
\colhead{$\mu_\Omega$} & 
\colhead{$\mu_\mathrm{los}$} & 
\colhead{$\Gamma$} & 
\colhead{$f_\mathrm{pl}/f_\mathrm{unscatt}$\tablenotemark{a}} &
\colhead{$\omega_\mathrm{bb}^\mathrm{fit}/\omega_\mathrm{bb}^\mathrm{sim}$} &
\multicolumn{5}{c}{Pulsed fraction (\%)}\\
 & & & & & & & \colhead{I} & \colhead{II} & \colhead{III} & \colhead{IV} & \colhead{V}
}
\startdata
0.2  & 1   & 0            & 0             & 7.6\tablenotemark{b} & \nodata & 0.96 & 43 & 46 & 52 & 76 & 46 \\
0.3  &     &              &               & 5.5\tablenotemark{b} & \nodata & 0.92 & 40 & 45 & 57 & 81 & 47 \\   
0.4  &     &              &               & 4.3                  & \nodata & 0.88 & 41 & 48 & 64 & 81 & 50 \\
0.5  &     &              &               & 3.5                  & 23      & 0.90 & 42 & 49 & 67 & 83 & 52 \\
0.6  &     &              &               & 2.9                  & 8.2     & 0.92 & 42 & 48 & 72 & 83 & 52 \\
0.7  &     &              &               & 2.4                  & 3.9     & 0.94 & 42 & 47 & 75 & 86 & 51 \\
0.8  &     &              &               & 2.0                  & 2.0     & 0.95 & 43 & 46 & 80 & 91 & 50 \\
0.75 & 0.1 & 0            & 0             & 2.7                  & 0.3     & 0.99 & 15 & 18 & 64 & 92 & 23 \\
     & 0.3 &              &               & 2.5                  & 0.9     & 0.98 & 30 & 37 & 78 & 89 & 44 \\
     & 0.7 &              &               & 2.3                  & 2.1     & 0.96 & 40 & 46 & 78 & 88 & 53 \\
     & 1.0 &              &               & 2.2                  & 2.8     & 0.95 & 42 & 47 & 77 & 89 & 51 \\
     & 1.3 &              &               & 2.2                  & 3.3     & 0.94 & 43 & 44 & 78 & 89 & 49 \\
0.6  & 1   & 1            & 1             & 3.5                  & 1.2     & 0.96 & 0  & 0  & 0  & 0  & 0  \\
     &     &              & $1/\sqrt{2}$  & 3.4\tablenotemark{b} & 10      & 0.88 & 0  & 0  & 0  & 0  & 0  \\
     &     &              & 0             & 3.1\tablenotemark{b} & 22      & 0.89 & 0  & 0  & 0  & 0  & 0  \\
     &     &              & -$1/\sqrt{2}$ & 2.6                  & \nodata & 0.86 & 0  & 0  & 0  & 0  & 0  \\
     &     &              & -1            & 3.0                  & 8.9     & 0.94 & 0  & 0  & 0  & 0  & 0  \\
     &     & $1/\sqrt{2}$ & $1/\sqrt{2}$  & 3.2\tablenotemark{b} & 6.0     & 0.92 & 35 & 27 & 41 & 50 & 19 \\
     &     &              & 0             & 2.9                  & 22      & 0.88 & 10 & 27 & 59 & 71 & 39 \\
     &     & 0            & 1             & 3.1                  & 20      & 0.90 & 0  & 0  & 0  & 0  & 0  \\
0.75 & 1   & 1            & 1             & 2.6\tablenotemark{b} & 0.5     & 0.97 & 0  & 0  & 0  & 0  & 0  \\
     &     &              & $1/\sqrt{2}$  & 2.6                  & 2.3     & 0.92 & 0  & 0  & 0  & 0  & 0  \\
     &     &              & 0             & 2.3                  & 4.6     & 0.93 & 0  & 0  & 0  & 0  & 0  \\
     &     &              & -$1/\sqrt{2}$ & 2.0                  & 7.9     & 0.92 & 0  & 0  & 0  & 0  & 0  \\
     &     &              & -1            & 2.5                  & 4.6     & 0.95 & 0  & 0  & 0  & 0  & 0  \\
     &     & $1/\sqrt{2}$ & $1/\sqrt{2}$  & 2.5                  & 2.1     & 0.94 & 37 & 25 & 46 & 66 & 13 \\
     &     &              & 0             & 2.2                  & 4.2     & 0.93 & 13 & 27 & 64 & 79 & 43 \\
     &     & 0            & 1             & 2.3                  & 4.5     & 0.93 & 0  & 0  & 0  & 0  & 0  \\
\enddata

\tablenotetext{a}{Entries with no data mean that setting $f_\mathrm{unscatt} = 0$ in eqn~(\ref{eq:fit})
provided the best fit.}
\tablenotetext{b}{Poor fit. See section~\ref{s:fit} for details on the accuracy of the fitting procedure.}
\end{deluxetable}


\begin{deluxetable}{cccclccccccc}
\tablecaption{Spectral Fit Parameters and Pulsed Fractions for Type II
Velocity Distribution (Boltzmann, bi-directional),
$\Delta\beta = 0.1$\label{t:pars_bimax}}
\tablewidth{0pt}
\tablehead{
\colhead{$\beta_0$} & 
\colhead{$\Delta \phi_\mathrm{N-S}$} & 
\colhead{$\mu_\Omega$} & 
\colhead{$\mu_\mathrm{los}$} & 
\colhead{$\Gamma$} & 
\colhead{$f_\mathrm{pl}/f_\mathrm{unscatt}$\tablenotemark{a}} &
\colhead{$\omega_\mathrm{bb}^\mathrm{fit}/\omega_\mathrm{bb}^\mathrm{sim}$} &
\multicolumn{5}{c}{Pulsed fraction (\%)}\\
 & & & & & & & \colhead{I} & \colhead{II} & \colhead{III} & \colhead{IV} & \colhead{V}
}
\startdata
0.2  & 1   & 0            & 0             & 7.7\tablenotemark{b} & \nodata & 0.99 & 35 & 28 & 12 & 33 & 27 \\
0.3  &     &              &               & 5.6\tablenotemark{b} & \nodata & 0.95 & 35 & 28 &  8 & 19 & 25 \\
0.4  &     &              &               & 4.3\tablenotemark{b} & \nodata & 0.91 & 37 & 28 &  5 & 16 & 24 \\
0.5  &     &              &               & 3.6                  & \nodata & 0.87 & 39 & 30 &  7 & 20 & 24 \\
0.6  &     &              &               & 3.0                  & 11      & 0.92 & 40 & 32 & 11 & 24 & 24 \\
0.7  &     &              &               & 2.5                  & 4.8     & 0.94 & 42 & 34 & 16 & 30 & 24 \\
0.8  &     &              &               & 2.1                  & 2.4     & 0.96 & 44 & 36 & 20 & 40 & 24 \\
0.75 & 0.1 & 0            & 0             & 2.7                  & 0.3     & 0.99 & 15 & 16 & 33 & 44 & 18 \\
     & 0.3 &              &               & 2.5                  & 0.9     & 0.98 & 30 & 32 & 38 & 36 & 33 \\
     & 0.7 &              &               & 2.4                  & 2.4     & 0.96 & 40 & 37 & 25 & 22 & 32 \\
     & 1.0 &              &               & 2.3                  & 3.4     & 0.95 & 43 & 34 & 18 & 34 & 24 \\
     & 1.3 &              &               & 2.2                  & 4.2     & 0.95 & 45 & 30 & 27 & 44 & 14 \\
0.6  & 1   & 1            & 1             & 3.3                  & 4.9     & 0.96 &  0 &  0 &  0 &  0 &  0 \\
     &     &              & $1/\sqrt{2}$  & 2.9                  & \nodata & 0.84 &  0 &  0 &  0 &  0 &  0 \\
     &     &              & 0             & 3.0                  & \nodata & 0.85 &  0 &  0 &  0 &  0 &  0 \\
     &     & $1/\sqrt{2}$ & $1/\sqrt{2}$  & 3.0                  & 14      & 0.92 & 40 & 31 & 11 & 20 & 24 \\
     &     &              & 0             & 2.9                  & \nodata & 0.84 &  6 &  6 &  9 & 11 &  7 \\
     &     & 0            & 1             & 3.0                  & \nodata & 0.85 &  0 &  0 &  0 &  0 &  0 \\
0.75 & 1   & 1            & 1             & 2.6                  & 2.6     & 0.97 &  0 &  0 &  0 &  0 &  0 \\
     &     &              & $1/\sqrt{2}$  & 2.1                  & 4.5     & 0.93 &  0 &  0 &  0 &  0 &  0 \\
     &     &              & 0             & 2.3                  & 5.8     & 0.93 &  0 &  0 &  0 &  0 &  0 \\
     &     & $1/\sqrt{2}$ & $1/\sqrt{2}$  & 2.3                  & 3.8     & 0.95 & 43 & 34 & 18 & 32 & 24 \\
     &     &              & 0             & 2.2                  & 5.3     & 0.93 & 13 & 12 & 16 & 15 & 14 \\
     &     & 0            & 1             & 2.3                  & 5.8     & 0.93 &  0 &  0 &  0 &  0 &  0 \\
\enddata
\tablenotetext{a}{Entries with no data mean that setting $f_\mathrm{unscatt} = 0$ in eqn~(\ref{eq:fit})
provided the best fit.}
\tablenotetext{b}{Poor fit. See section~\ref{s:fit} for details on the accuracy of the fitting procedure.}
\end{deluxetable}

\begin{deluxetable}{cccccclccccccc}
\tablecaption{Spectral Fit Parameters and Pulsed Fractions for Type III Velocity
Distribution \\ (Broad, Relativistic) \label{t:pars_broad}}
\tablewidth{0pt}
\tablehead{
\colhead{$\beta_\mathrm{min}$} &
\colhead{$\gamma_\mathrm{max}$} &
\colhead{$\alpha$} &
\colhead{$\Delta \phi_\mathrm{N-S}$} &
\colhead{$\mu_\Omega$} &
\colhead{$\mu_\mathrm{los}$} &
\colhead{$\Gamma$} & 
\colhead{$f_\mathrm{pl}/f_\mathrm{unscatt}$\tablenotemark{a}} &
\colhead{$\omega_\mathrm{bb}^\mathrm{fit}/\omega_\mathrm{bb}^\mathrm{sim}$} &
\multicolumn{5}{c}{Pulsed fraction (\%)}\\
 & & & & & & & & & \colhead{I} & \colhead{II} & \colhead{III} & \colhead{IV} & \colhead{V}
}
\startdata
0.1 & 3    & -1 & 1   & 0             & 0             & 2.0                  & 1.3     & 0.96 & 39 & 43 & 79 &  90 & 48 \\
0.3 &      &    &     &               &               & 1.8                  & 1.1     & 0.96 & 38 & 38 & 83 &  92 & 45 \\
0.7 &      &    &     &               &               & 1.5                  & 0.6     & 0.96 & 36 & 34 & 89 &  94 & 52 \\
0.9 &      &    &     &               &               & 1.1                  & 0.3     & 0.98 & 36 & 34 & 92 &  96 & 54 \\
0.3 & 1.1  & -1 & 1   & 0             & 0             & 4.0\tablenotemark{b} & \nodata & 0.83 & 34 & 39 & 40 &  45 & 38 \\
    & 1.96 &    &     &               &               & 2.3                  & 2.4     & 0.93 & 38 & 39 & 75 &  81 & 42 \\
    & 3.48 &    &     &               &               & 1.7                  & 0.9     & 0.97 & 38 & 37 & 84 &  94 & 45 \\
    & 6.19 &    &     &               &               & 1.5                  & 0.5     & 0.98 & 38 & 36 & 85 &  97 & 43 \\
    & 11   &    &     &               &               & 1.4                  & 0.4     & 0.98 & 38 & 36 & 84 &  98 & 39 \\
    & 19.6 &    &     &               &               & 1.4                  & 0.3     & 0.99 & 38 & 36 & 83 &  99 & 35 \\
    & 34.8 &    &     &               &               & 1.4                  & 0.2     & 0.99 & 38 & 36 & 81 & 100 & 32 \\
0.3 & 3    & -4 & 1   & 0             & 0             & 3.3\tablenotemark{b} & 8.9     & 0.90 & 36 & 41 & 55 &  69 & 42 \\
    &      & -3 &     &               &               & 2.8\tablenotemark{b} & 4.2     & 0.92 & 37 & 41 & 63 &  78 & 42 \\
    &      & -2 &     &               &               & 2.3                  & 2.0     & 0.95 & 38 & 40 & 74 &  86 & 42 \\
    &      &  0 &     &               &               & 1.5                  & 0.7     & 0.97 & 37 & 35 & 87 &  94 & 50 \\
0.3 & 3    & -1 & 0.1 & 0             & 0             & 2.1                  & 0.1     & 1.00 &  6 &  9 & 73 &  95 & 18 \\
    &      &    & 0.3 &               &               & 2.0                  & 0.4     & 0.99 & 15 & 22 & 83 &  93 & 37 \\
    &      &    & 0.7 &               &               & 1.9                  & 0.9     & 0.97 & 30 & 35 & 82 &  91 & 44 \\
    &      &    & 1.3 &               &               & 1.8                  & 1.2     & 0.95 & 43 & 37 & 84 &  92 & 46 \\
0.3 & 3    & -1 & 1   & 1             & 1             & 2.4\tablenotemark{b} & 0.3     & 0.96 &  0 &  0 &  0 &   0 &  0 \\
    &      &    &     &               & $1/\sqrt{2}$  & 2.0                  & 0.5     & 0.95 &  0 &  0 &  0 &   0 &  0 \\
    &      &    &     &               & 0             & 1.6                  & 1.0     & 0.96 &  0 &  0 &  0 &   0 &  0 \\
    &      &    &     &               & -$1/\sqrt{2}$ & 1.7                  & 2.4     & 0.95 &  0 &  0 &  0 &   0 &  0 \\
    &      &    &     &               & -1            & 2.3                  & 3.5     & 0.94 &  0 &  0 &  0 &   0 &  0 \\
    &      &    &     & $1/\sqrt{2}$  & $1/\sqrt{2}$  & 1.8                  & 0.5     & 0.96 & 29 & 17 & 61 &  82 &  9 \\
    &      &    &     &               &  0            & 1.7                  & 1.1     & 0.96 & 28 & 23 & 65 &  80 & 38 \\
    &      &    &     & 0             &  1            & 1.6                  & 1.0     & 0.96 &  0 &  0 &  0 &   0 &  0 \\
\enddata

\tablenotetext{a}{Entries with no data mean that setting $f_\mathrm{unscatt} = 0$ in eqn~(\ref{eq:fit})
provided the best fit.}
\tablenotetext{b}{Poor fit. See section~\ref{s:fit} for details on the accuracy of the fitting procedure.}
\end{deluxetable}

A more distinct high-energy component of the spectrum is created when the charges have
a broad, relativistic distribution of momenta (Fig.~\ref{f:wFw_broad}).  The energy spectrum 
dips above the thermal peak and then rises as a power-law up to a maximum frequency
$\gamma_{\rm max}^2\,\omega_{\rm peak}$.  The photon index above the dip
can be related directly to the index $\alpha$ of the particle distribution by assuming
that each high-energy photon has scattered only once.  The number flux of photons upscattered
to a frequency $\gamma^2\omega_{\rm peak}$ is ${1\over\hbar}F_\omega \propto \gamma f
\propto \gamma^{1+\alpha} \propto \omega^{(1+\alpha)/2}$.  The high-energy photon index is
therefore
\begin{equation}
\Gamma = {1-\alpha\over 2}.
\end{equation}
A flat particle number distribution ($\alpha = -1$) results in a rising energy spectrum,
$\omega F_\omega \propto \omega$; whereas a flat energy distribution ($\alpha = -2$)
results in an energy spectrum that is somewhat softer, $\omega F_\omega \propto \omega^{1/2}$.

Geometric effects also strongly influence the spectrum that the observer
sees.  The highly anisotropic character of the emission pattern
can be appreciated by comparing the spectra of the aligned rotator
($\mu_\Omega = 1$: Figs~\ref{f:wFw_sim}c,e; \ref{f:wFw_e+e-}c,e; and 
Fig.~\ref{f:wFw_broad}g) with those of the orthogonal rotator 
($\mu_\Omega = 0$: Figs~\ref{f:wFw_sim}d,f; \ref{f:wFw_e+e-}d,f; 
and Fig.~\ref{f:wFw_broad}h).  Changes in the observer's orientation (the value of
$\mu_{\rm los}$) manifest themselves through variations in the relative flux 
of the non-thermal component, and also through changes of the order of
$\Delta\Gamma \sim 1$ in the photon index.  Except for the nearly-aligned
rotator, the differences in spectral hardness resulting from a
change in the observer's orientation are smaller than those which divide the
spectrum of a quiescent AXP from the spectrum of an SGR.

\subsection{Pulse Profiles and Pulsed Fractions}\label{s:pulse_prof}

The persistent emission of the SGRs and AXPs is pulsed.
The pulse profiles generally have one or two main pulses in the 1-10 keV band,
but range from simple sinusoids to more complex, multiple-peaked profiles.
The pulsed fraction, defined as
\begin{equation}\label{eq:pulsedfrac}
PF = {F_{\rm max}-F_{\rm min}\over F_{\rm max}+F_{\rm min}}
\end{equation}
where $F_{\rm min,max}$ are the minimum and maximum fluxes measured
within one rotation period,  ranges from a few percent to as high 
as $\sim 70\%$ \citep{wt06}.  A complex pulse profile is seen more frequently
in the persistent emission of the SGRs.

\begin{figure}
\vskip .4in
\plotone{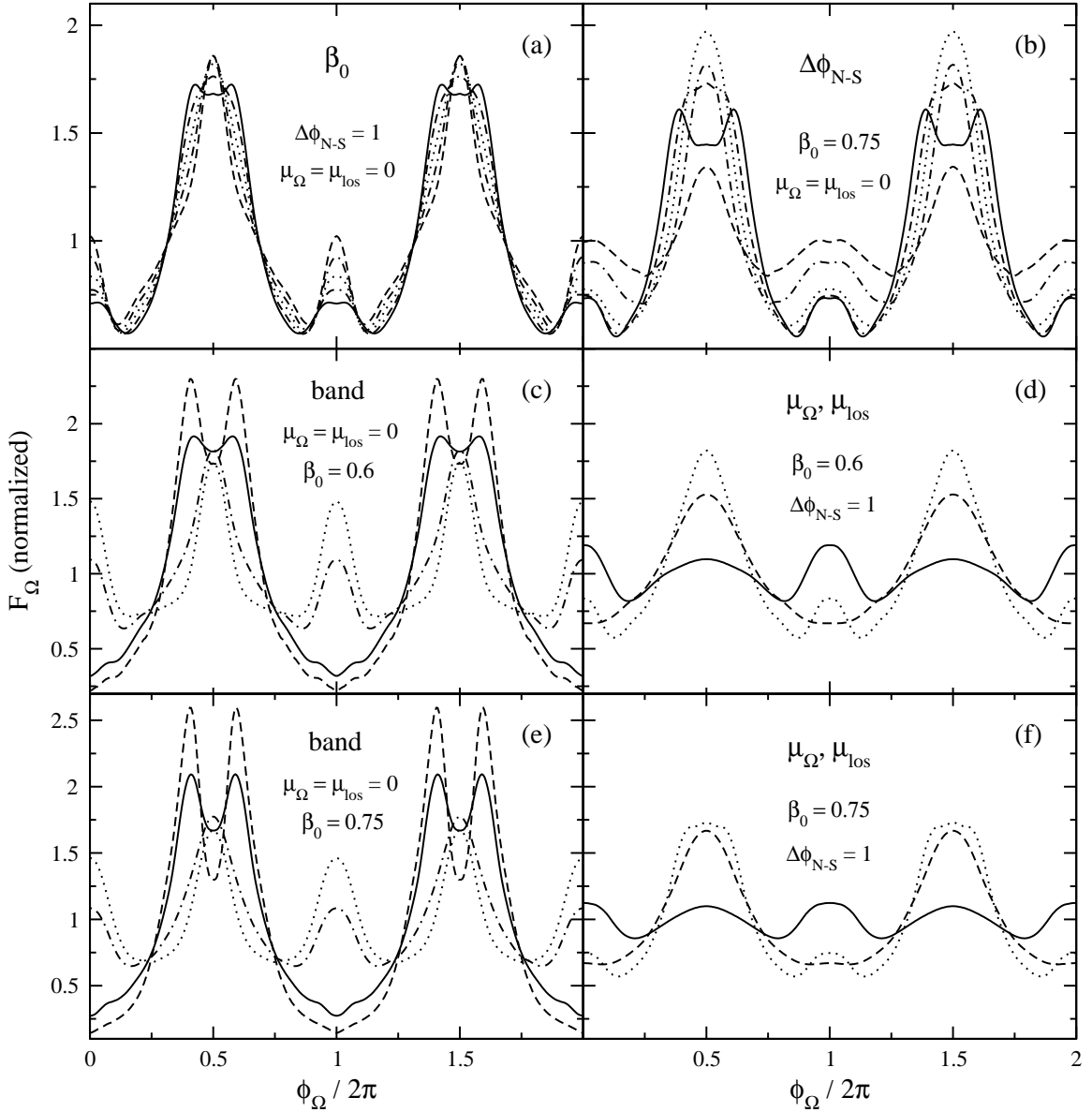}
\caption{Pulse profiles obtained with a unidirectional Boltzmann particle distribution 
(type I; eq.~[\ref{eq:boltzone}]).  All curves are normalized so that the average amplitude
is unity. Simulation parameters of (a), (b), (c)-(d), (e)-(f) are the same as in the corresponding panels in 
Fig.~\ref{f:wFw_sim}.  Energy bands are defined in eq.~(\ref{eq:energy_bands}).
Curves in panel (a) show the band V profile for different velocity spreads:
$\beta_0 = \{0.4$ (dash), $0.5$ (dot-dash), $0.6$ (dotted), $0.7$ (dash-dash-dot), $0.8$ (solid)$\}$.
Band V pulse profiles in panel (b) correspond to different twist angles:  
$\Delta \phi_\mathrm{N-S} = \{0.1$ (dash), $0.3$ (dot-dash), $0.7$ (dotted), $1$ (dash-dash-dotted), 
$1.3$ (solid)$\}$.  Curves in panels (c) and (e) are for different bands:  I (dotted), 
II (dot-dash), III (solid), and IV (dash), with $\beta_0 = 0.6$ and $0.75$, respectively. 
Band V pulse profiles in panels (d) and (f) correspond to different orientations of the magnetic axis and the line
of sight with respect to the spin axis:  $(\mu_\Omega,\mu_\mathrm{los}) = \{(1/\sqrt{2},1/\sqrt{2})$ (solid),
$(1/\sqrt{2},0)$ (dash), $(0,0)$ (dotted)$\}$, with $\beta_0 = 0.6$ and $0.75$, respectively. 
Pulsed fractions in bands I-V for all configurations are listed in Table~\ref{t:pars_max}.
}
\label{f:pulse_sim}
\end{figure}

\begin{figure}
\vskip .4in
\plotone{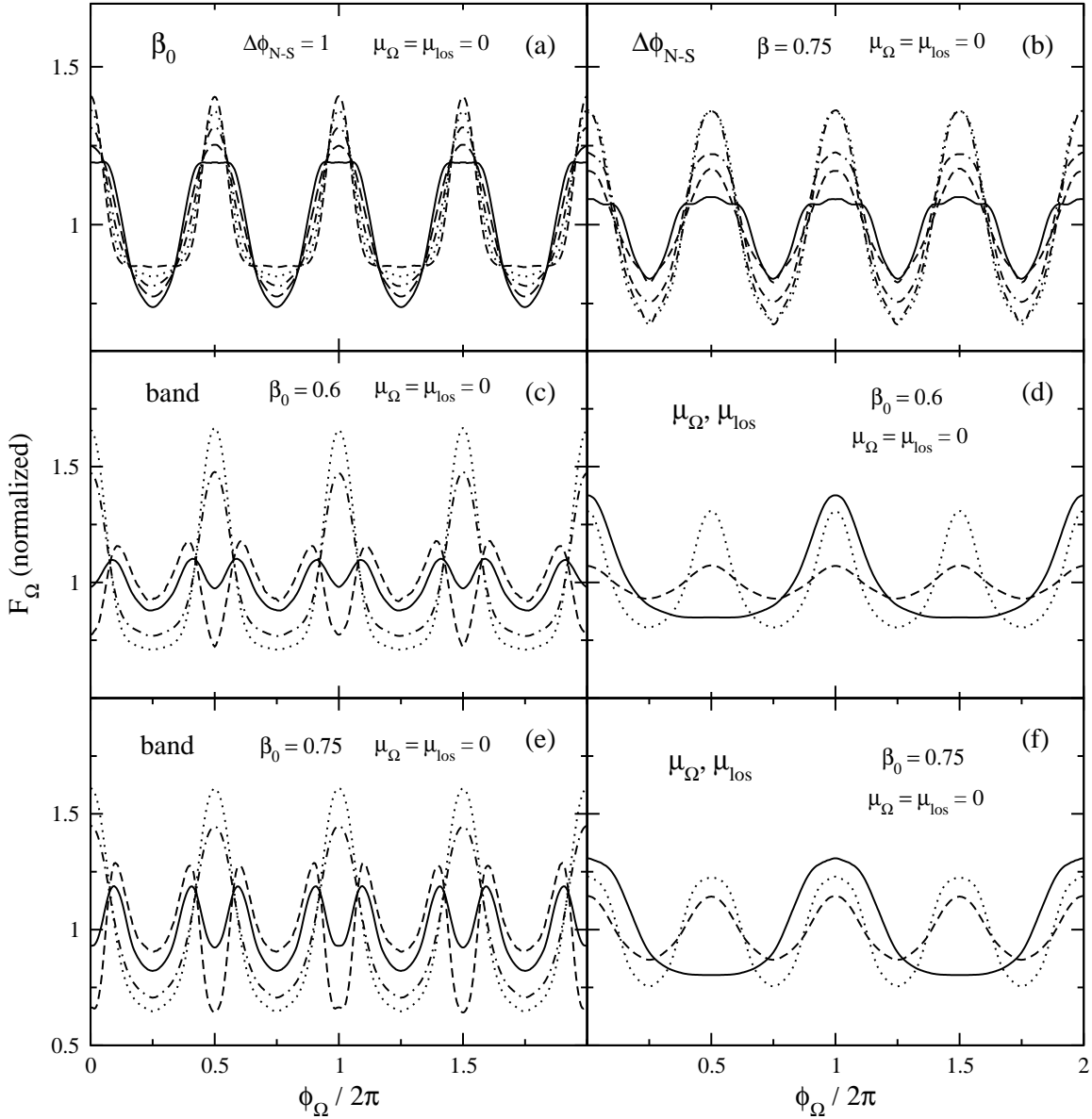}
\caption{Same as Fig.~\ref{f:wFw_sim}, but for a velocity
distribution function that is bi-directional,
corresponding to a pair plasma (type II; eq.~[\ref{eq:boltzone}]). 
Simulation parameters of (a), (b), (c)-(d), (e)-(f) are the same
as those in the corresponding panels of Fig.~\ref{f:wFw_e+e-}. 
Pulsed fractions are listed in Table~\ref{t:pars_bimax}.}
\label{f:pulse_e+e-}
\end{figure}

\begin{figure}
\vskip .4in
\plotone{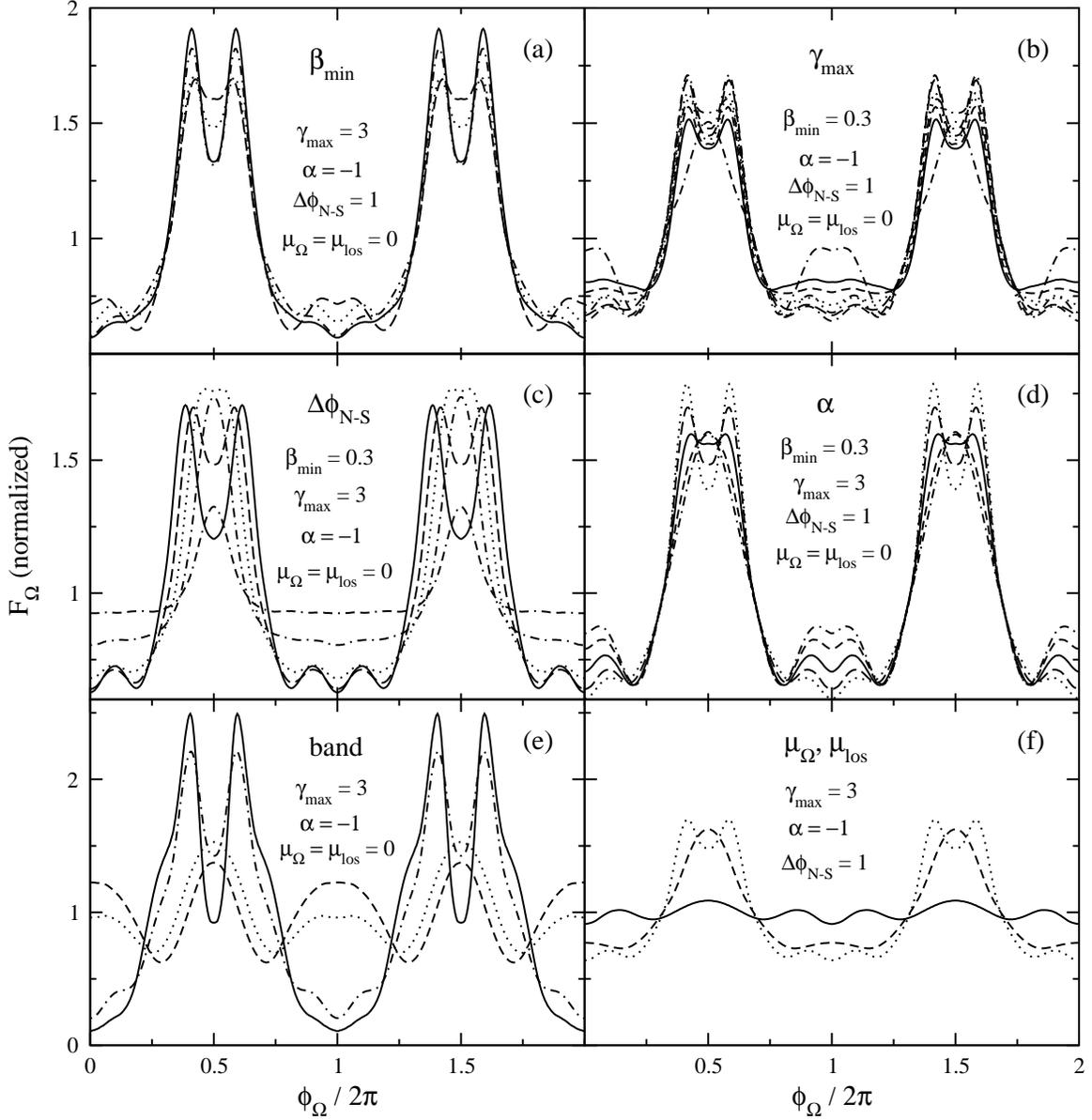}
\caption{
Pulse profiles obtained with the broad velocity distribution function, eq.~(\ref{eq:broad_dist}),
with same normalization as Fig.~\ref{f:pulse_sim}. Simulations parameters are as shown 
on each panel. 
Panel (a) shows profiles in band V obtained with different minimum velocities, with curves corresponding
to $\beta_\mathrm{min} = \{0.1$ (dash), $0.3$ (dotted), $0.7$ (dot-dash), $0.9$ (solid)$\}$. 
Panel (b) shows the result of varying the maximum Lorentz factor, with curves corresponding to
$\gamma_\mathrm{max} = \{1.1$ (dash-dash-dot), $1.96$ (dot-dot-dash), $3.48$ (dash-dot), $6.19$ (long dash),
$11$ (dotted), $19.6$ (short dash), $34.8$ (solid)$\}$. Panel (c) shows results for different twist angles, 
with curves corresponding to $\Delta \phi_\mathrm{N-S} = \{0.1$ (dash-dash-dot), $0.3$ (dash-dot), $0.7$ (dotted),
$1$ (long dash), $1.3$ (solid)$\}$. Panel (d) shows the result of varying the exponent of the distribution
function, with curves corresponding to $\alpha = \{-4$ (dot-dash), $-3$ (short dash), $-2$ (solid), $-1$ 
(long dash), $0$ (dotted)$\}$. Panel (e) shows lighcurves in bands I (dash),
II (dotted), III (dot-dash), and IV (solid), for $\beta_\mathrm{min}= 0.3$, $\gamma_\mathrm{max} = 3$, $\alpha = -1$,
$\Delta \phi_\mathrm{N-S} = 1$, and $\mu_\Omega = \mu_\mathrm{los} = 0$. Panel (f) shows profiles in band V
obtained from the same simulation as in (e), but varying the orientation as 
$(\mu_\Omega,\mu_\mathrm{los}) = \{ (1/\sqrt{2},1/\sqrt{2})$ (solid line), $(1/\sqrt{2},0)$ (dashed line),
$(0,0)$ (dotted line)$\}$. Pulsed fractions are shown in Table~\ref{t:pars_broad}.
}
\label{f:pulse_broad}
\end{figure}

The pulse profiles resulting from particle velocity distributions of types
I-III are displayed in Figs.~\ref{f:pulse_sim}-\ref{f:pulse_broad}.
(These figures corrrespond to the spectra in Figs. \ref{f:wFw_sim}-\ref{f:wFw_broad}.)
The profiles are plotted in five frequency bands, which are taken to be
\begin{equation}
\label{eq:energy_bands}
\begin{array}{ccccccc}
\textrm{I}  & : &   &      &\omega/\omega_\mathrm{peak} & \leq & 1\\
\textrm{II} & : & 1 & \leq &\omega/\omega_\mathrm{peak} & \leq & 3\\
\textrm{III}& : & 3 & \leq &\omega/\omega_\mathrm{peak} & \leq & 10\\
\textrm{IV} & : &   &      &\omega/\omega_\mathrm{peak} & \geq & 10\\
\textrm{V}  & : & 1 & \leq &\omega/\omega_\mathrm{peak} & \leq & 10,
\end{array}
\end{equation} 
where $\omega_\mathrm{peak}$ is defined in eq.~(\ref{eq:omega_peak}).
Each pulse profile is normalized by the pulse-averaged
flux in that band:  averaging over the rotation gives unit flux.
The pulsed fractions are listed band by band in Tables~\ref{t:pars_max}-\ref{t:pars_broad}; the values are generally accurate to 1\%. 

The pulse profiles have a wide variety of morphologies, and sometimes vary significantly with 
frequency.  Several properties can be understood from the fact that the
direction of a scattered photon is collimated more tightly about the direction
of the magnetic field as the speed of the scattering charge increases.
The differential cross section for resonant scattering is proportional to
$(1+{\mu'}^2)$, where $\mu'$ is the direction cosine in the rest frame of the charge after
scattering (e.g. \citealt{meszaros92}). The probability of scattering at an 
angle $\mu > 0 $ in the stellar frame is $P(\mu>0) = (4 + 3\beta + \beta^3)/8$ [see \S~\ref{s:angles}].
For $\beta = 0.5$, there is a 70\% probability of photons being scattered in the same
direction as the charge is moving.  

Several trends are apparent in Figs.~\ref{f:pulse_sim}-\ref{f:pulse_broad}.
The pulsed fraction increases with particle drift
speed, as does the asymmetry between the main pulse and
the sub-pulse 180 degrees away in phase (Fig.~\ref{f:pulse_sim}a).  
The main pulse is emitted
toward the south magnetic pole, and the sub-pulse toward the north magnetic pole. 
Although the scattering particles flow out of the north pole and into the south pole,
the scattering depth is largest near the magnetic equator, where the particles flow
southward.   
The asymmetry between main pulse and the sub-pulse also grows with
increasing twist,
because of the growing asymmetry in the photon flux near the magnetic equator due to
the increased optical depth there (Fig.~\ref{f:pulse_sim}b).  At the same time the main pulse divides
into two sub-pulses, because the density of charge carriers vanishes along the
magnetic axis.  The sub-pulse is generally more distinct in the bands near the peak
of the input thermal spectrum, because these photons are less scattered that those
at higher energies (Figs.~\ref{f:pulse_sim}c,e).  Finally, the pulsed fraction is
generally larger for nearly orthogonal rotators (Figs.~\ref{f:pulse_sim}d,f).

In spite of this diversity, the axisymmetry of the magnetic field translates into
a reflection symmetry of the pulse profile about phase $\phi_\Omega = \pi$.
The absence of such a mirror symmetry is a direct signature of some non-axisymmetric
structure in the surface heat flux or in the magnetospheric currents, which is not accounted for in the present study.

The pulse profile also provides information about the composition of the magnetospheric
corona.  Some of the trends in Fig.~\ref{f:pulse_sim} are preserved when the
velocity distribution is of type II (symmetric under $\beta \rightarrow -\beta$),
but some differences emerge.  Figure~\ref{f:pulse_e+e-} displays the resulting pulse profiles;
all other parameters are held constant from Fig.~\ref{f:pulse_sim}.  When the magnetic axis is 
nearly orthogonal to the rotation axis (as in panels a,b,c and e), the observer must
see two comparable pulses, because the charges stream in both directions along
the magnetic field.  For an orthogonal rotator, as displayed here, the light curve 
between rotational phase $180^\circ$ and $360^\circ$ is, in fact, a mirror image of the 
light curve between $0^\circ$ and $180^\circ$; but this orientation is a special case.
More generally is possible for the observer to see a single pulse, as demonstrated in 
Fig.~\ref{f:pulse_e+e-}d,f in the case where $\mu_\Omega = 1/\sqrt{2}$.   

A single pulse can divide into two sub-pulses for the same reason as in Fig.~\ref{f:pulse_sim}:
the current density and scattering rate decreases toward the magnetic axis.  One interesting
feature of Fig.~\ref{f:pulse_e+e-} is that the positions of the sub-pulses can alternate
by $\pi/2$ in phase between different bands:  this reflects the number of scatterings
that a photon in each band has typically experienced.  

It should be emphasized that, aside from the number of major sub-pulses, which depends directly on the
composition of the magnetospheric plasma, most properties of the pulse profiles depend even more strongly
on orientation.   A good example is provided by Fig.~\ref{f:pulse_sim}f, which shows how
a single-peaked profile with subpulses (dotted line) can  be transformed into a profile without any
subpulse (dashed line), and then to a double-peaked profile (solid line) merely by changing
$\mu_\Omega$ and $\mu_{\rm los}$.

The calculated pulsed fractions $P$ range from 0\% to nearly 100\%. 
The magnitude of $P$ and its dependence on the simulation parameters
can be quite different in the low-energy and high-energy bands.
In general, the pulsed fraction correlates positively with twist angle 
as long as $\Delta \phi_\mathrm{N-S}\lesssim 1$.
The largest pulsed fractions are obtained for the broad velocity distribution (eq. [\ref{eq:broad_dist}]), as is shown in Fig.~\ref{f:pulse_broad}.  

It should be kept in mind that the heat flux is expected to vary strongly over the surface if the elastic
deformations of the crust are localized in a small area:  the small radiative
areas of SGR and AXP afterglows (about 1 percent of the neutron star surface
area: \citealt{ibrahim01}; \citealt{wt06}) provide evidence for this behavior.
Our assumption of a single-temperature blackbody therefore involves
a considerable simplification of reality, at least in the more transient
magnetar sources.  The combined effects of anisotropic emission and 
magnetospheric scattering are not examined here.

\subsection{Spectral Features}

\citet{lyutikov06} suggest that spectral features in the 
seed photon frequency distribution may be smeared out as the result
of multiple scattering in a magnetar magnetosphere. 
Here we test this effect with our three-dimensional Monte Carlo model.

\begin{figure}
\vskip .4in
\epsscale{1.05}
\plottwo{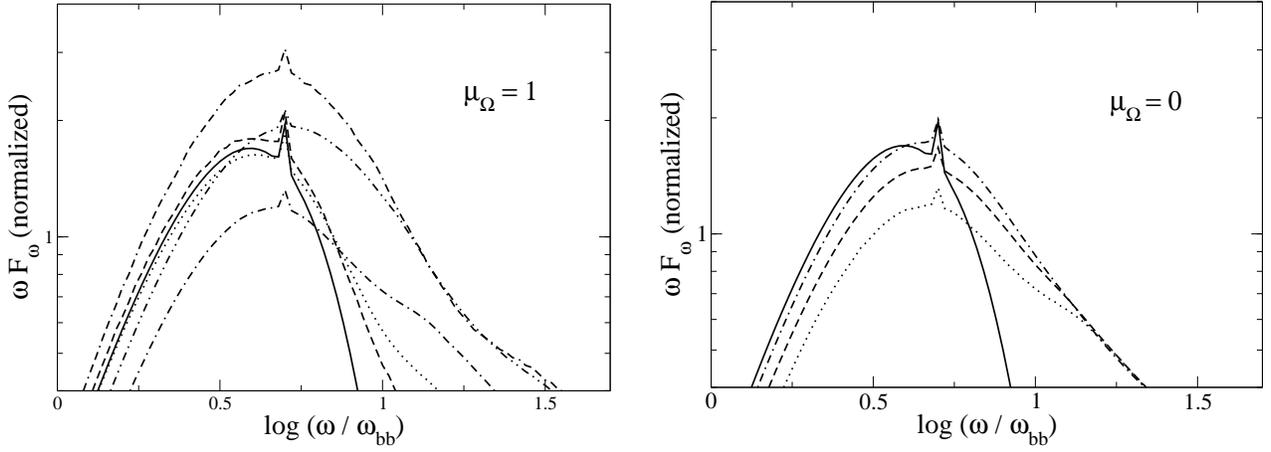}{f10b.eps}
\caption{Effect of magnetospheric scattering on an emission line.
Input spectrum (solid curves) has a line centered at 
$\omega = 5\,k_{\rm B}T_{\rm bb}/\hbar$, with half-width 
$0.1\,k_{\rm B}T_{\rm bb}/\hbar$
and total energy equal to 1\% of the blackbody.  Output spectra are obtained 
by convolving the input with the magnetospheric response function for 
different orientations.  \emph{Left:} Aligned rotator ($\mu_\Omega = 1$), 
with $\mu_\mathrm{los} = 1$ (dash), $1/\sqrt{2}$ (dotted), $0$ (dot-dash), $-1/\sqrt{2}$ (dot-dot-dash),
and $-1$ (dash-dash-dotted). \emph{Right:} Orthogonal rotator ($\mu_\Omega = 0$), with 
$\mu_\mathrm{los} = 1$ (dotted), $1/\sqrt{2}$ (dash), and $0$ (dot-dash).
Velocity distribution is mono-energetic and undirectional 
(eq.~[\ref{eq:tophat}]), with a mean drift speed $\bar \beta = 0.5$
and a total width $\Delta \beta = 0.1$.  The twist angle is
$\Delta \phi_\mathrm{N-S} = 1$.}

\label{f:wFw_line}
\end{figure}

Figure~\ref{f:wFw_line} shows energy spectra obtained with a narrow 
velocity distribution ($\bar \beta = 0.5$, $\Delta \beta = 0.1$) and 
large twist ($\Delta \phi_{\rm N-S} = 1$).   Instead of convolving the
response function with a plain black body spectrum, we now add an emission
line to the seed photon distribution. The line is centered at 
$\omega = 5k_{\rm B}T_{\rm bb}/\hbar$, 
with half-width $0.1\omega_\mathrm{bb}$, 
and total flux 1\% of the integrated blackbody flux, as done by 
\citet{lyutikov06}. The spectra observed from both an aligned rotator 
and an orthogonal rotator are plotted.  Even though the line strength is 
reduced by a factor $\sim 2$ in some orientations, the line is not smeared 
out.  The basic reasons are that i) the optical depth is never very large
along any direction; and ii) a substantial fraction (of the order of
one half) of the emitted black body photons must escape the magnetosphere 
directly without scattering, even if the magnetospheric twist is strong. 
The greatest amount of smearing would be expected in a geometry where a 
significant optical depth is maintained along the line of sight over an 
entire rotation of the star.  Nonetheless, a significant line feature
remains even in the case of a nearly aligned rotator, with the line of 
sight nearly orthogonal to the magnetic and rotation axes 
(the lowest curve in Fig.~\ref{f:wFw_line}a).

\subsection{Ion Scattering}
\label{s:ionscatt}

We have performed some preliminary calculations of the effects of ion 
cyclotron scattering on the output spectrum.   We consider the case of
protons (with a charge/mass ratio $1/1836$ that of the electron).  A polar
magnetic field of $1\times 10^{15}$ G is chosen, which allows
proton cyclotron scattering over an order-of-magnitude range in frequency
above the black body peak.

\begin{figure}
\vskip .4in
\epsscale{0.95}
\plotone{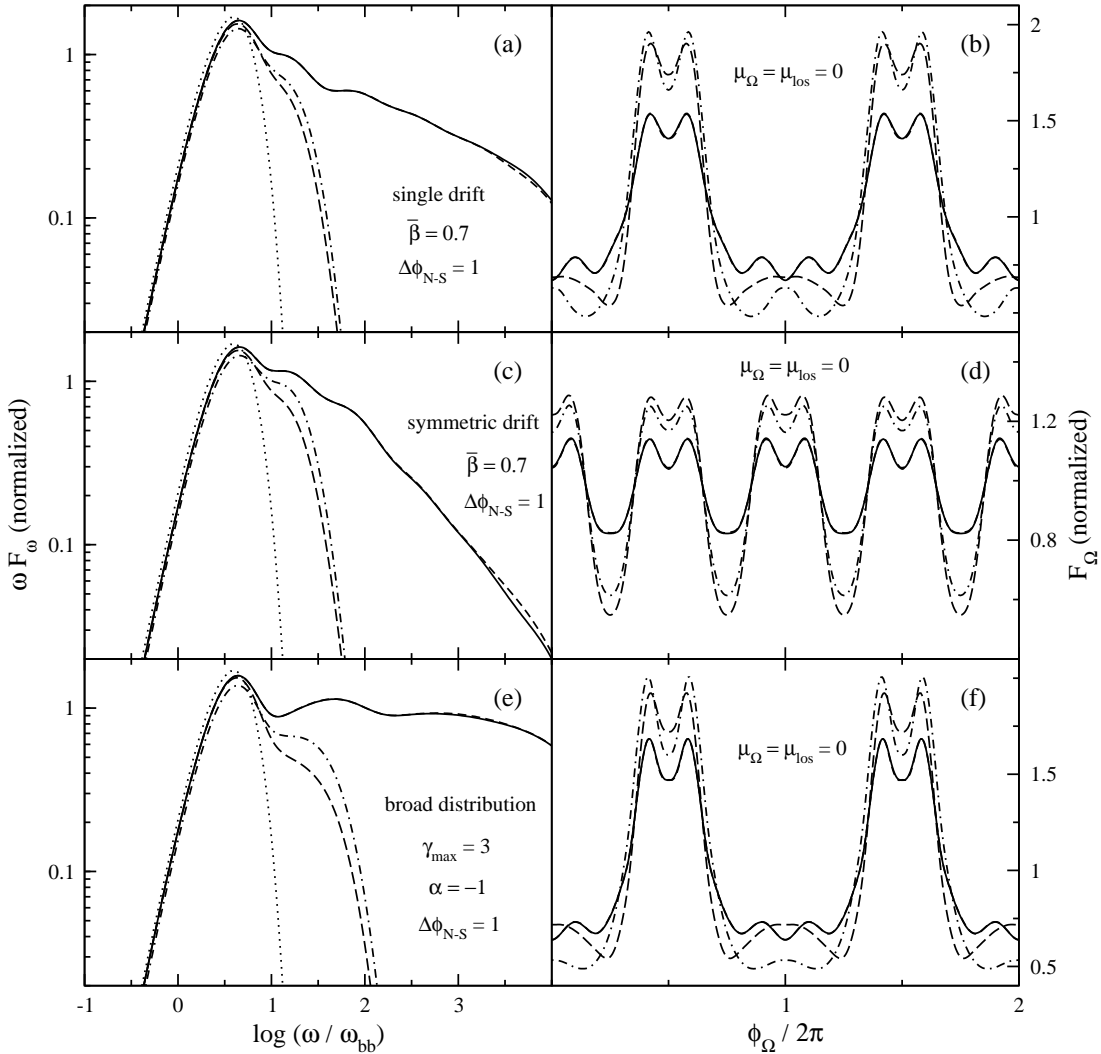}
\caption{Energy spectra (left) and band V pulse profiles (right) 
obtained for $B_{\rm pole} = 10^{15}$ G and a charge/mass
ratio equal to that of the proton and the
electron, for radial (long-dash: proton, solid: electron) 
and isotropic (dot-dashed: proton, short-dashed: electron) 
angular distribution of the seed photons.   
Panels (a) and (b) are for a mono-energetic,
unidirectional particle distribution (eq.~[\ref{eq:tophat}]) with
mean drift $\bar \beta = 0.7$ and $\Delta \beta = 0.1$.
In panels (c) and (d), the velocity distribution is mono-energetic
and bi-directional, as is appropriate to the case where the ions
are driven off the surface of the magnetar by radiation pressure.
Other parameters are the same.
In panels (e) and (f), the velocity distribution
is broad and relativistic (type III; eq.~[\ref{eq:broad_dist}]), 
with $\beta_\mathrm{min} = 0.3$, $\gamma_\mathrm{max}=3$ and $\alpha = -1$.
All curves assume $\Delta \phi_\mathrm{N-S}=1$,
and $\mu_\Omega = \mu_\mathrm{los} = 0$. Curves for electrons are almost
undistinguishable.} 
\label{f:ion}
\end{figure}

Energy spectra and pulse profiles are shown in Fig.~\ref{f:ion}.  The results for protons
are plotted side by side with those for electrons, using the same 
position-independent 
velocity 
distribution (eqs. \ref{eq:boltzone}-\ref{eq:broad_dist}), and varying the angular distributions 
of the seed photons. Solid lines and
dot-dashed lines show results for protons assuming radial and isotropic seed photon
emission, respectively, with solid and short-dashed curves containing correponding results 
for electrons. The first two sets of figures assume a narrow top hat velocity distribution.  
The case of a bi-directional proton velocity
distribution can be related to a situation in which the X-ray flux at
the proton cyclotron line exceeds the Eddington flux, and a cloud of ions and electrons is
blown off the stellar surface and suspended in the magnetosphere \citep{tlk02}. 
The final set of figures assume a broad particle momentum distribution (type III; 
eq.~[\ref{eq:broad_dist}]).

Multiple ion scattering generates a high-frequency power law tail to 
the input thermal spectrum.  The spectrum is however softer than that
produced by electron scattering:  photons that are backscattered by ions to the star can be absorbed
at the stellar surface (with unit probability in the E-mode, and with probability $\sim {1\over 2}$ in 
the O-mode).  As expected, the high-energy spectral tail is
cut off sharply at a frequency $(r_\mathrm{bb}/R_\mathrm{NS})^{(2+p)}\omega_{\rm peak} \sim 10\,
\omega_{\rm peak}$.  In this calculation, the cutoff frequency for electron scattering
is $m_p/m_e\sim 10^3$ times higher.  We have not displayed this cutoff because
the calculation does not include the effects of electron recoil (which becomes
important at 50-100 keV), or relativistic corrections to the scattering 
cross section at the first Landau resonance.  
Recoil effects can be entirely neglected for ion scattering.

The effects of ion scattering could in principle be discerned during a very luminous phase, e.g.
the afterglow following an intermediate-energy SGR burst.  For example, 
the afterglow of the 29 August 1998 burst of SGR 1900$+$14 showed a marked spectral hardening in the 
2-5 and 5-10 keV bands, but not in the 10-20 keV band \citep{ibrahim01}. 

Results for proton scattering are sensitive to the angular distribution of seed photons, given
their smaller resonant radius (Fig.~\ref{f:ion}). 
Their spectra become harder when the angular distribution is
changed from radial to isotropic. In contrast, spectra obtained with electron scattering are
almost independent of the input angular distribution, with results for the radial and
isotropic case being hardly distinguishable.

The pulses produced by ion and electron scattering have a similar morphology (Fig.~\ref{f:ion}),
although electron scattering yields smaller pulsed fractions. Again, results for protons are
somewhat sensitive to the angular distribution of seed photons, while the same parameter does not
influence pulse profiles obtained with electron scattering. A proper account of ion scattering 
must include the effects of light bending of the photon trajectories 

\subsection{Seed Photon Polarization}
\label{s:polarization}

There are two possible sources of seed X-ray photons:  internal heating
by a decaying magnetic field (e.g. \citealt{td96,hk98,arras04}), and
the bombardment of the surface by energetic charges \citep{tlk02}.
In the first case, the seed photons are emitted in the E-mode if
the magnetic field is stronger than $\sim 10^{14}$ G \citep{holai04}.  
In the second
case, the emission is in the O-mode if the charges are stopped by
Coulomb collisions in a surface layer that is optically thick to 
free-free absorption.

It is therefore interesting to explore whether these two types of
seed photons can be distinguished using the output X-ray spectrum or
pulse profiles.  The scattering cross sections of the two modes at
the cyclotron resonance are comparable, because the electromagnetic
eigenmodes are determined by vacuum polarization effects and are
linearly polarized even in the core of the cyclotron resonance
(\S \ref{s:rescyc}).  For this reason, one expects similar pulse
profiles and non-thermal spectra for both polarizations 
of the seed photons.  Nonetheless, the scattering cross section of
an O-mode photon is suppressed with respect to that of an E-mode photon
by a factor $(\mu')^2$ (eqs. [\ref{eq:sigma_resb}], [\ref{eq:circular_amplitude}];
see also the discussion in Appendix~\ref{s:tests}).
One therefore expects the spectra to be slightly softer and the pulsed
fractions to be slightly lower when the seed photons are injected in
the O-mode.

\begin{figure}
\vskip .4in
\plottwo{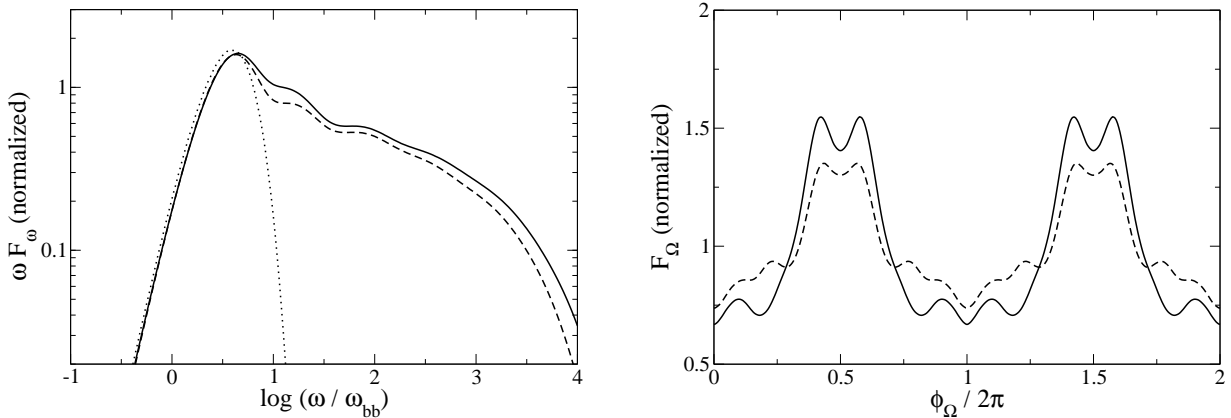}{f12b.eps}
\caption{
Energy spectra (left) and band V pulse profile (right) corresponding
to 100\% polarization of the seed photons in the E-mode (solid curves)
and the O-mode (dashed curves).  Velocity distribution is mono-energetic and
and unidirectional (eq. [\ref{eq:tophat}]) with mean drift speed 
$\bar \beta = 0.7$ and total width $\Delta \beta = 0.1$.  In addition, 
$\Delta \phi_\mathrm{N-S} = 1$, and $\mu_\Omega = \mu_\mathrm{los} = 0$.
}
\label{f:EOmode}
\end{figure}

Figure~\ref{f:EOmode} shows energy spectra (left) and pulse profiles in band V (right)
using the same magnetospheric parameters, excepting that the seed photons are 
100\% polarized in the E-mode (solid lines) and the O-mode (dashed lines).
In the second case, the non-thermal tail does indeed have a slightly lower
amplitude compared with the black body peak, and the pulsed fraction is
smaller.
Nevertheless, the changes in the emission pattern resulting from the
polarization of the seed photons are much subtler than those produced
by changes in the orientation of the star 
(\S \ref{s:pindex}, \ref{s:pulse_prof}), 
offering little chance of identifying the seed photon emission mechanism
except through direct polarization measurements.

\section{Discussion}

We have developed a Monte Carlo code to study the reprocessing
of thermal X-rays by resonant cyclotron scattering in neutron star 
magnetospheres using the model of \citet{tlk02}. The code is fully 
three dimensional, and allows for an arbitrary velocity 
distribution of the scattering particles and an arbitrary magnetic 
field geometry.  Angle-dependent X-ray spectra and pulse profiles
can be calculated for an arbitrary orientation between the 
magnetic axis and the spin axis, and between the spin axis and the
line of sight to the star.

Our aim here has been to show which types of X-ray spectra
and pulse profiles follow generically from the model, and which 
may point to different emission mechanisms.  We can reproduce
most of the properties of the 1-10 keV emission of the AXPs,
but not of the SGRs, by assuming a broad and mildly relativistic
velocity distribution, and moderate twist angles
($\Delta\phi_{\rm N-S} \sim 0.3-1$ radian).   In particular,
the SGRs in their most active states have harder X-ray spectra
than can be reproduced by the model \citep{woods06}.
Our results demonstrate
that changes in the strength of the twist and in the dynamics of the 
charged particles (which are coupled strongly to the radiation field)
can lead to substantial changes in flux, hardness, pulse shape, and 
pulsed fraction.  

\subsection{Relative Strength of the Thermal and Non-thermal Components
   of the X-ray Spectrum}
\label{s:sumone}

The X-ray spectra of most magnetars display a prominent
thermal component.  The corresponding blackbody area is
less than, or sometimes comparable to, the expected surface
area of a neutron star (e.g. \citealt{wt06}).   In sources
where the blackbody component is dominant at $\sim 1-3$ keV, the high-energy
photon index is typically $2$ or softer.  SGR 1900$+$14 presents an
example of an SGR where a thermal component and a hard power-law
component of the spectrum are simultaneously present \citep{esposito06}.

Model spectra with a mildly relativistic particle distribution
($k_{\rm B}T_0 \simeq 0.5\,m_ec^2, \beta_0 \simeq 0.75$; 
eq. [\ref{eq:boltzone}]) can easily reproduce the 1-10 keV 
spectra of sources like 4U 0142$+$614, 1RXS J170849$-$4009 and 1E 1841$-$045.
Some of our model spectra show a noticeable downward break at 
$\sim 30$-$50\,k_{\rm B}T_{\rm bb} \simeq 10$-30 keV
(e.g. Figs. 4f, 5f).  Such a break may be present in the
combined XMM and Integral spectra of SGR 1900$+$14
\citep{gotz06}, and its presence is not excluded in other
sources such as the AXPs 4U 0142$+$614 and 1E 1841$-$045
when one takes into account the presence of a separate
rising high-energy spectral component above 20 keV
\citep{kuiper06, gotz06}.  

The quiescent SGRs have relatively weak
blackbody components to their spectra, and during periods
of extreme activity a pure power-law fit to the spectrum
can have a photon index of $\sim -1.6$ \citep{mereghetti05,
woods06}.  (Even harder photon indices are obtained from
combined blackbody-powerlaw fits to the spectrum, but then the
continuation of the 2-10 keV fit to higher energies
disagrees with the Integral flux measurements.)
We could obtain spectra as hard as this by choosing a broad
power-law momentum distribution for the magnetospheric charges
(Fig. 6), but the output spectrum then displays a strong blackbody
component.  In the fit of \citet{esposito06} for SGR 1900$+$14,
the amplitude of the black body component is about three times
that of the power-law component at the black body peak, whereas in the spectra
of Fig. 6b, it is 6-8 times larger.

We expect this conclusion to hold for any
model for the high-energy continuum which relies on the upscattering
of the thermal seed photons.  Balancing the energy input to the
coronal charges with the loss by upscattering, one finds
that $\tau (\langle\gamma^2\rangle-1) \sim 1$ (as in corona models based
on non-magnetic Compton scattering).  When $\langle\gamma^2\rangle^{1/2} 
\gg 1$, it is generally not possible to upscatter a large fraction of the 
thermal photons because the optical depth $\tau$ of the corona
is small.   The results obtained with a broad relativistic
particle distribution function should also be treated with caution:
the charges are subject to a strong 
cyclotron drag force at a radius of $\sim 50$-100 km where the cyclotron
energy is 1-10 keV, and so it is probably not self-consistent to
assume that they have the same distribution function there as they
do in the inner parts of the magnetosphere.

These considerations point to the conclusion that the hard non-thermal
spectral states of the SGRs (with photon indices $< 2$) have
a similar origin to the {\it excess} high energy emission that is
observed from several AXPs above 20 keV.  In the magnetar model,
this high-energy emission must involve a different emission
process, such as bremsstrahlung from a hot transition layer between
the relativistic corona and the colder neutron star atmosphere;
or synchrotron emission triggered by pair cascades at a
distance of $\sim 100$ km, where the electron cyclotron energy is
$\sim 1$ keV \citep{tb05}.  

Indeed, there are reasons to expect that some
active magnetars (e.g. the SGRs) can have significantly reduced
X-ray emission due to deep internal cooling, even when the time-averaged
rate of magnetic field decay remains high.   Large increases in the neutrino 
emissivity are expected when the core neutrons undergo a superfluid 
transition, which leads to an order-of-magnitude drop in blackbody
X-ray flux \citep{arras04}.  The greater intermittency of the magnetic
dissipation in a Soft Gamma Repeater (as compared with the AXPs) could
then be tied to the greater rigidity of the neutron star crust at lower
temperatures.
Given the high absorbing columns $N_H \sim 2-6\times 10^{22}$
cm$^{-2}$ that are inferred from the soft X-ray spectra of the two
active SGRs, it would then be possible to hide the surface thermal
component from detection.
(The combination of a low thermal luminosity and strong absorption 
does not remove the inconsistency with the model spectra in Fig. 6c-e: 
the models then would predict that the non-thermal continuum is much 
dimmer than $\sim 10^{35}$ ergs s$^{-1}$ at a photon energy of a few keV.)

\subsection{Low Energy X-ray Spectra}\label{s:lowen}

Our results show that the rescattering of thermal photons by magnetospheric
charges does not lead to any enhanced emission at energies
below the peak of the seed X-ray distribution. 
Our new fitting formula, eq.~(\ref{eq:fit}), takes this low energy behavior of
resonant scattering into account.  

It should be kept in mind that the softest power-law components in measured 
spectra are typically needed to fit a {\it low-energy} excess to a
single-temperature black body.  
In our model most of the emission below the thermal peak
comes from unscattered photons: our rotationally-averaged
spectra have a Rayleigh-Jeans shape at low energies, which lies just
below the input blackbody distribution.  We therefore suggest
that a fit to an AXP X-ray spectrum involving a small $N_H$ and a very soft
power law component is physically inconsistent, and that in such a case
the fitting formula (\ref{eq:fit}) is strongly preferred.
Of course, the softest AXP spectra can also be fitted by 
a superposition of two or more black body components
(e.g. \citealt{halpern05}).

The formation of a high-energy tail to the black-body spectrum is closely
tied, in this model, to an increased spindown rate (due to the flaring
out of the poloidal field lines as they are twisted).  The
AXP 1E 2259$+$586 provides an example of a magnetar which has demonstrated
SGR-like burst activity \citep{woods04}, but which also has the lowest
spindown rate of any magnetar candidate, as well as very soft
1-10 keV spectrum, and only an upper bound on the 100 keV emission that
has been detected from several other AXPs \citep{kuiper06}.  We would
therefore expect that the 1-10 keV spectrum of 1E 2259$+$586 is most
accurately modelled by a variable surface temperature, rather than
by a combination of thermal and non-thermal components.   

Indeed, a recent reanalysis of the X-ray spectra of several AXPs has revealed
a broader spectrum than a pure single-temperature black body near the thermal 
peak \citep{durant06a}.  This suggests that the thermal X-ray flux is 
distributed inhomogeneously across the surface of the neutron star.  This 
is expected given the inhomogeneous nature of the yielding behavior in the
crust of a magnetar, and the anisotropy of the thermal conductivity in 
the presence of a strong magnetic field \citep{td96,let02,perez06,geppert05}.

\subsection{Pulse Profiles and Pulsed Fractions}\label{s:sumtwo}

Our calculations generally reproduce X-ray pulse profiles that are
characteristic of the AXPs: single or multiple peaks, with sinusoidal 
shapes and sometimes the presence of narrow sub-pulses.  
The strong angular dependence of the optical depth to resonant scattering
explains why phase resolved spectra show variations both in their flux level,
as well as in hardness (e.g. \citealt{tiengo05,goehler05}).  
We find that the X-ray spectrum is generally hardest at pulse maximum,
in agreement with some measurements \citep{mereghetti04, tiengo05}.

The number of sub-pulses depends essentially on the orientation of the spin and
magnetic axes of the neutron star, which are not constrained by 
any independent measurements.  The pulse profiles of the actively
bursting SGRs tend to be more complicated, which may reflect
presence of high-order multipoles in the current flowing near the
stellar surface.  As discussed in \S \ref{s:sumone}, their
very hard X-ray spectra are more consistent with a separate
emission process, possibly localized close to the surface of the neutron
star.  

Pulsed fractions as high as $\sim 40-50\%$ near the black body peak
are easily obtained from the anisotropic scattering of thermal photons
in the magnetosphere (Tables 1-3). Significantly higher pulsed fractions
are possible in the power-law tail of the spectrum, due to the
increased anisotropy of the multiply-scattered photons.  It should
be emphasized that this result is obtained assuming a {\it spherically
symmetric} source of seed photons, i.e. a constant temperature 
across the surface of the neutron star.  Only in one object
(the AXP 1E 1048.1$-$5957) is the pulsed fraction significantly higher 
(70-80\%) in the 1-10~keV range \citep{tiengo05}.  It should be possible
to obtain higher pulsed fractions if the seed photons are emitted
at localized hotspots, as suggested by the small black body emitting
area \citep{durant06b}.   Our calculations are therefore complementary
to those of \citet{pod00}, \citet{opk01}, and \citet{ozel02}, who
studied the pulse profiles and pulsed fractions 
that result from isolated hotspots on rotating
neutron stars, and compared their calculations with the observed 
1-10 keV pulse profiles of the AXPs.

\subsection{Comparison with Previous Work}

Semi-analytical work on resonant cyclotron scattering in the atmospheres of neutron stars goes back to the 1980s, 
the original focus being the transfer of radiation through the atmosphere of the star, or through an accretion
column \citep{yahel79,wasserman80,nagel80,meszaros85a,meszaros85b}.  In these calculations, the geometrical thickness of the
scattering material was generally taken to be small compared with the gradient scale of the magnetic field.
The effect of magnetic field inhomogeneity was analyzed by \citet{zheleznyakov96}, who found solutions to the
one-dimensional radiative transfer equation using the Schwarshild-Shuster method. \citet{lyutikov06} generalized
this method to include the combined effects of the particle motion and magnetic field inhomogeneity.  In the limit of
small particle dispersion $\beta c$, they found that the radiation spectrum that is transferred through an optically thick
medium is upscattered in energy by a factor $1+2\beta$.  This model was shown to provide a good fit to the
1-10 keV spectrum of 1E 1048.1$-$5937, the AXP source in which the thermal component of the spectrum peaks
at the highest frequency.  However, this non-relativistic and one-dimensional approximation cannot provide
a realistic description of the more extended high-energy spectra that are observed in some magnetars, including
the slope of a power-law tail to the input thermal spectrum, the relative amplitude of the power-law and thermal
components, or the detailed shape and frequency-dependence of the pulse profiles.

Monte Carlo codes for resonant cyclotron scattering through the surface layers of a neutron star 
were developed around the same time (e.g. \citealt{pravdo81,wang88}), and also applied to the problem
of line formation in gamma-ray bursts \citep{wang93}.  The last two studies remain the
most realistic to date, as they include the effects of thermal broadening,  
vacuum and plasma polarization, as well as electron recoil.  All of these
effects but the last one are included in our study (plasma makes a negligible
contribution to the dielectric properties of the magnetosphere at frequencies
of $10^{17}-10^{19}$ Hz).  However, these studies are all one-dimensional and
focus on the regime of very large optical depth and a weak gradient in the
magnetic field, and so the results cannot be compared directly with ours.

\subsection{Further Developments}

We now discuss how some of the key simplifications of the model will
need to be relaxed in future work.

\begin{enumerate}
\label{s:radpres}
\item \emph{Anisotropic radiation pressure.} The most important simplification
in the treatment of the transfer of radiation through the magnetosphere is the
decoupling of the particle dynamics from the radiation force that is imparted
through cyclotron scattering.  An electric field is induced by the twisted 
magnetic field lines, which lifts charged particles off the stellar surface,
and accelerates them to high speeds (\citealt{tlk02}, \citealt{bt06}).
Where the Lorentz-boosted cyclotron energy lies in the range 1-10 keV
(or 1-100 keV for an active SGR), the particle moving against the direction
in which the radiation is flowing feel a strong drag force, and a strong
electric field is required to compensate \citep{tb05}.  The charged
particle dynamics and the radiation transfer are therefore strongly coupled.

\item\emph{Particle recoil.}
The non-inclusion of particle recoil after scattering leads to significant
errors in the shape of the output spectra above $\sim 100$ keV.  
The spectra obtained with broad relativistic particle distributions
involve essentially the single scattering of photons of a frequency
$\sim \omega_{\rm peak} = 2.82\,k_{\rm B}T_{\rm bb}/\hbar$.  
Recoil is important only when $\gamma \ga m_ec^2/2\hbar\omega_{\rm peak}$ and
$\hbar\omega \sim \gamma^2\omega_{\rm peak} \ga 10$ MeV, where we do not
tabulate spectra.  It should be emphasized that this
remains a technical issue insofar as additional emission mechanisms
dominate at $\sim 30$-100 keV (see \S \ref{s:sumone}).

\item \emph{Seed photon distribution.} In all our simulations we have 
assumed that the seed photons are emitted radially, from points uniformly 
distributed over the stellar surface, and with a single polarization. 
There are several reasons for this choice.  First, the radius of first
scattering by electrons (or positrons) is $R_{\rm res} \sim 10\,R_{\rm NS}$;
second, we are not including the effects temperature inhomogeneities on
the neutron star surface, focusing instead on in the pulsations that result
from the transport of X-ray photons through a twisted magnetosphere.
The effects of light bending must be included as soon as this
second assumption is relaxed.

\item \emph{Polarization.}   Continuum spectra and pulsed profiles do not
provide a clear way of discriminating between thermal emission from the 
surface due to deep cooling (E-mode emission) and due to particle
bombardment and stopping by Coulomb collisions (O-mode emission).  
In both cases, the surface temperature will in practice be anisotropic
(due to the channeling of the cooling heat flux along the magnetic field,
and due to variations in the strength of the twist across the surface).
In the second case, one expects the intensity of the O-mode photons
to be more strongly beamed along the local direction of the surface
magnetic field (e.g. \citealt{basko75}), but we have found that pulsed
fractions of 40-50\% percent are achievable solely by radiation transport 
through the magnetosphere.  Polarization measurements of AXPs
and SGRs would say much about the physics of the beam-heated atmosphere
of a magnetar, and the distribution of magnetospheric currents, but they
require polarimeters that are sensitive at flux levels of $\sim 10^{-11}$
ergs cm$^{-2}$ s$^{-1}$ in the 1-10 keV band.  

\item \emph{Magnetic field geometry.}  The magnetospheric model
employed by \citet{tlk02} is self-similar:  the pitch angle of the external
field lines is independent of distance from the neutron star.
Clearly different X-ray spectra and pulse profiles will result if
the current is distributed in a different way throughout the magnetosphere.
Observations which show a delayed increase in the spindown torque following
periods of X-ray burst activity (\citealt{woods02}) suggest that the twist
is not distributed homogeneously throughout the magnetosphere in
the immediate aftermath of a crustal yielding event.
Quadrupole and octopole components of the magnetic field have a large
effect on the pulse profile only if ions are the dominant scattering species,
or if the non-thermal emission is localized close to the neutron star
surface (as it is with bremsstrahlung emission; \citealt{tb05}).

\end{enumerate}

\acknowledgments

We thank Andrei Beloborodov, Maxim Lyutikov, Kaya Mori, Martin Durant, 
Marten van Kerkwijk, and Chris Matzner for useful discussions.  
Comments by an anonymous referee helped to clarify the presentation 
of our results. 
This work was supported by NSERC of Canada.  Parallel computations 
were performed on CITA's McKenzie cluster \citep{dubinski03} which 
was funded by the Canada Foundation for Innovation and the Ontario 
Innovation Trust.

\begin{appendix}

\section{A. Code Tests}
\label{s:tests}

\subsection{Input Frequency Distribution and General Output}
\label{s:response_test}

\begin{figure}
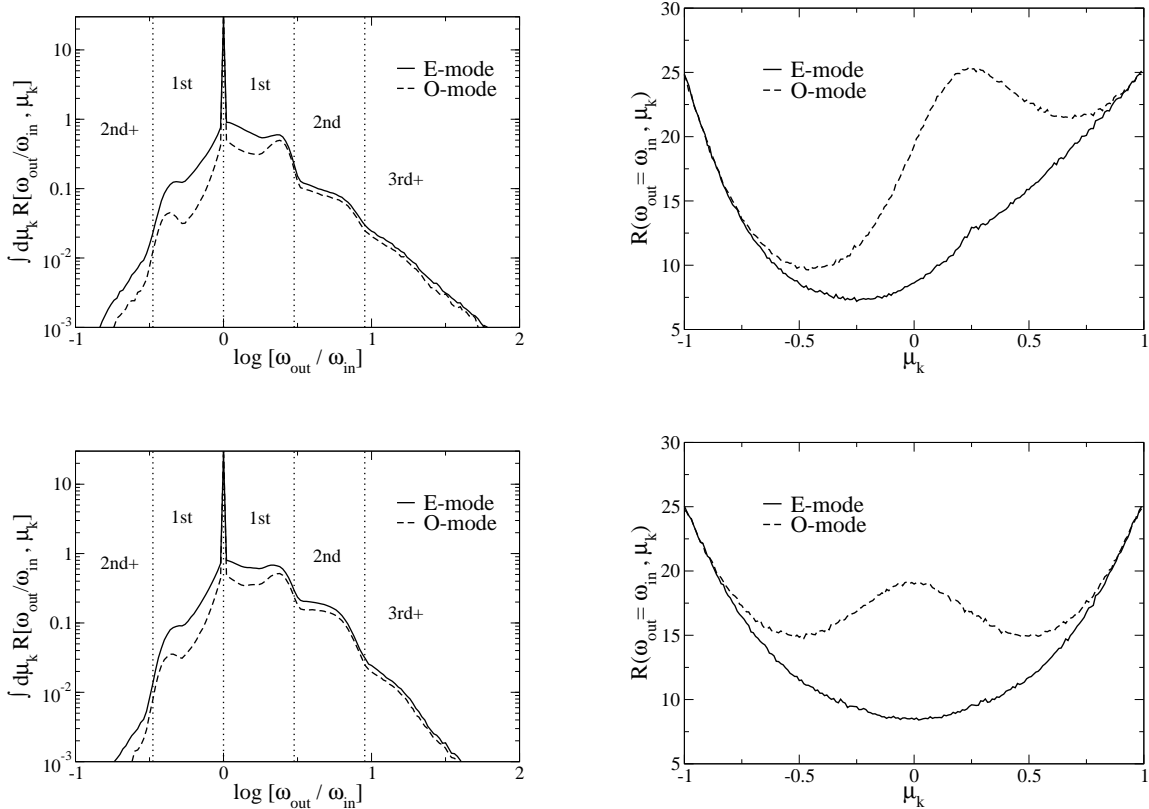

\vskip .4in
\epsscale{0.9}
\plottwo{f13a.eps}{f13b.eps}
\vskip .3in
\plottwo{f13c.eps}{f13d.eps}
\caption{\emph{Left:}  Output spectrum in a twisted magnetosphere 
($\Delta\phi_{\rm N-S} = 1$) obtained by integrating eq.~(\ref{eq:response}) 
over magnetic polar angle.  
Seed photons are injected in either the E-mode (solid lines) 
or the O-mode (dashed lines).  Vertical dotted lines mark out
the maximum frequency shift in each order of scattering; they
are separated by factors of $(1+|\beta|)/(1-|\beta|)$ in frequency.
\emph{Right:} Angular distribution of the
unscattered photons ($\omega_{\rm out} = \omega_{\rm in}$) with respect to
the magnetic polar axis.   In both cases the top plots assume a
mono-energetic particle distribution (eq. [\ref{eq:tophat}]) 
with mean drift speed $\bar\beta = 0.5$, total width $\Delta\beta = 0.1$, and
unidirectional particle flow.   The bottom plots assume the same 
distribution function, but a bidirectional particle flow.}
\label{f:response}
\end{figure}

The output spectrum resulting from the input of a monochromatic line 
(frequency $\omega_{\rm in}$)
is obtained by integrating the response function $R$ (eq.~[\ref{eq:response}])
over magnetic polar angle $\mu_\mathrm{k}$.  
Well sampled spectra and pulse profiles are obtained 
with $10^7$ photons per run, using
bins of size 0.02 in $\log(\omega)$ and 0.01 in $\mu_{\rm k}$.
The left two panels of
Fig.~\ref{f:response} show the result for a mono-energetic particle
distribution function (eq. [\ref{eq:tophat}]) with $\bar\beta = 0.5$, $\Delta \beta = 0.1$, and 
particle flow in either one or both directions along the magnetic field.
The two curves in each graph represent the injection of the seed photons
in the two polarization eigenmodes.

The output spectrum has a sharp peak at $\omega_\mathrm{out}
=\omega_\mathrm{in}$.  The fraction of the seed photons that escape
unscattered is shown in the right panel of Fig.~\ref{f:response}.
It is higher for the O-mode than for the E-mode, because the
O-mode cross section has an additional factor of $(\mu')^2$,
where $\mu' = \hat k\cdot\hat B$ is evaluated in the rest frame
of the charge (eqs. [\ref{eq:sigma_resb}], [\ref{eq:circular_amplitude}]).
The fraction of unscattered photons is highest along the two magnetic
poles, where the optical depth goes to zero.  Here it is independent of 
the polarization of the seed photons, because the two eigenmodes 
are degenerate when a photon propagates nearly parallel to ${\bf B}$.

The fraction of unscattered O-mode photons shows a third peak, which
arises because the cyclotron optical depth vanishes where $\mu = \bar\beta$.
This condition is equivalent to $\mu' = 0$:  the photon propagates 
orthogonally to the magnetic field in the rest frame of the charge.
(It is satisfied at $\mu_\mathrm{k} \simeq 2.77$ for radially moving photons
in a dipole magnetic field, where the particle drift is in only one
direction along the field at a speed $\bar\beta = 0.5$.  The small 
bump in the E-mode curve at this angle is a numerical artifact.)

Above the input frequency, the escaping photons have an approximately power law
distribution.   The power-law tail is present because i) the optical depth 
$\tau$ through
the corona is independent of resonant frequency (eq. [\ref{eq:taures}]), and
ii) because the particle distribution function is assumed to be independent
of position.  It follows from i) that each photon has a probability 
\begin{equation}\label{eq:Pofn}
P(n) = \left[1-\exp(-\tau)\right]^n
\end{equation}
of undergoing $n$ scatterings.  At the same
time, ii) implies that the frequency increment $\Delta\omega$
due to backscattering is independent of the scattering order.  The
probability that a photon is upscattered from a frequency $\omega_{\rm in}$
to $\omega_n
= (1+\Delta\omega/\omega)^n\,\omega_{\rm in}$ is therefore
\begin{equation}
\label{eq:powerlaw}
P(\omega_n) = 
\left(\frac{\omega_n}{\omega_{\rm in}}\right)^
{\left\{\ln{[1-\exp{(-\tau)}]}/\ln{[1+\Delta\omega/\omega]}\right\}}.
\end{equation}
The power-law index is negative here because $\Delta\omega > 0$.
It is of the order of unity when the particle drift is mildly relativistic,
because $\Delta\omega/\omega \sim 1$.  A more precise analytic calculation
of the spectral slope is very difficult in practice, because
$\tau$ and $\Delta\omega/\omega$ depend on $\mu$ and $\beta$ in a non-linear 
manner. See \citet{pozdnyakov83} for a general review of the effects of multiple
non-resonant Compton scattering in X-ray spectra. 

The spectral signature of the first few orders of scattering is
apparent in Fig. \ref{f:response}.  The maximum frequency following the
first backscattering is given by $\omega_{\rm out}/\omega_{\rm in} = 
(1+|\bar \beta|)/(1-|\bar \beta|)$, and following the second by
$[(1+|\bar\beta|)/(1-|\bar \beta|)]^2$.  One also sees the effect of 
downscattering to a frequency 
$[(1-|\bar \beta|)/(1+|\bar \beta|)]\omega_{\rm in}$,
since some of the charges are flowing outward (in at least one 
hemisphere).  The low-frequency spectral 
slope is generally harder than the Rayleigh-Jeans value of 2, which means
that the output spectrum has a nearly Rayleigh-Jeans shape at low
frequencies when the input spectrum is a single-temperature black body.
The observation of a flatter spectrum at lower frequencies in some
AXP sources \citep{durant06a} therefore points to the presence of
temperature gradients across the neutron star surface.  

\begin{figure}
\vskip .4in
\plottwo{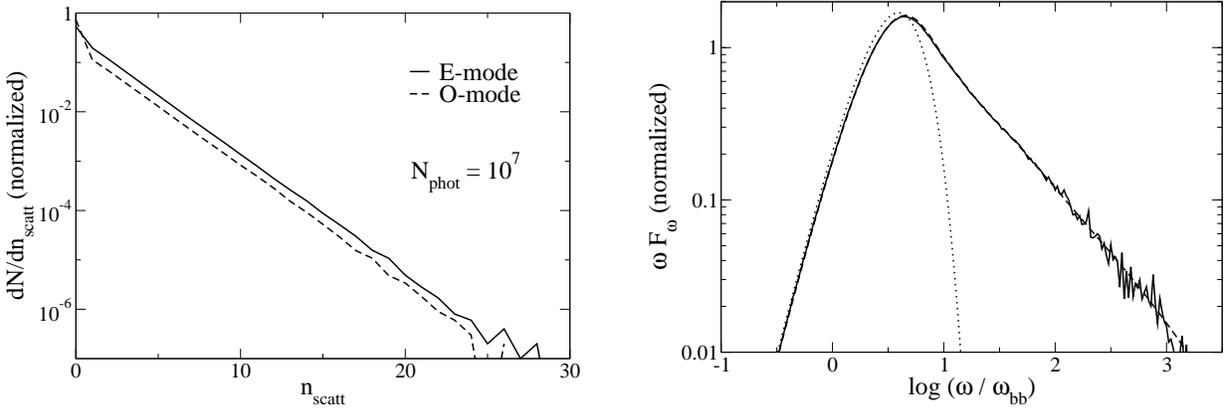}{f14b.eps}
\caption{\emph{Left:} Distribution of number of scatterings
for the simulation shown in the upper panels of Fig.~\ref{f:response}.
As expected in a multiple scattering process with optical depth independent
of the order of scattering (eq.~[\ref{eq:powerlaw}]), the distribution is 
exponential. Both seed polarizations generate the same slope, although 
the number of unscattered photons is bigger for the O-mode due to its
smaller mean optical depth.
\emph{Right:} Energy spectrum for the same simulation parameters and
E-mode seed photons, but with a different handling of the input frequency
distribution.  The solid line shows the result obtained by drawing 
frequencies directly from a blackbody distribution (eq.~[\ref{eq:bb_number}]),
whereas the dashed is obtained by convolving the response function of
a monochromatic seed spectrum (frequency $2.82\,k_{\rm B}T_{\rm bb}/\hbar$; 
left panel of Fig.~\ref{f:response}) with a blackbody distribution.  Energy
spectra are normalized so that the area of the Planck function is unity.}
\label{f:bb_input}
\end{figure}

The distribution of number of scatterings $n$ is shown in 
the left panel of Fig.~\ref{f:bb_input} for the case of undirectional
particle flow (and the same simulation parameters as in 
Fig.~\ref{f:response}).  Matching the power-law slope with
eq. \ref{eq:Pofn}, one infers a small average optical depth
per scattering, $\tau \simeq 0.02$, for $\Delta\phi_{\rm N-S} = 1$
and $\bar\beta = 0.5$.  Both seed photon polarizations 
produce nearly the same slope.  The simulation includes enough photons
($N = 10^7$) to obtain good statistics for the first $\sim 20-25$
scatterings.

The output energy spectrum resulting from the input of a single-temperature black body 
is plotted in the right panel of Fig.~\ref{f:bb_input}.  Here two methods are used, which give
nearly identical results.  The continuous line shows the result obtained by drawing the seed photon 
frequencies directly from a blackbody number distribution (eq.~[\ref{eq:bb_number}]);  the dashed line 
shows the result obtained by using a single input frequency ($\omega_\mathrm{peak}$, eq.~\ref{eq:omega_peak}), 
and then convolving the output spectrum with a blackbody.   The main advantage of the second method is that
the output spectrum is somewhat smoother, and we adopt it throughout this paper.

The energy spectra obtained with the current code were cross-checked with the output
from a Monte Carlo code written previously by one of the authors (CT),
which assumed a (split)-monopole magnetic field geometry, a mono-energetic particle spectrum,
and calculated the position of successive scattering surfaces by a different (geometric) method.   This previous code treated the transport of a photon within
a resonant surface in a one-dimensional approximation.
The two codes gave identical results for the high-energy power-law index, given the same
field geometry, angular distribution of optical depth ($\sin^2\theta$; eq. [\ref{eq:taures}]),
and particle drift speed.  The output spectrum differed only in some details in the shape of
the peaks at the first and second scattering orders, which were ascribable to the neglect of the
spatial curvature of the resonant surface in the 1-D code.   In particular, 
a photon that is scattered in a direction
tangent to the resonant surface sees an unrealistically high optical depth when the
surface curvature is neglected.   This resulted
in the presence of a spurious dip at a frequency $(\omega_\mathrm{out}/\omega_\mathrm{in}) = 1/(1-|\bar \beta|)$
in the 1-D output spectrum.

\subsection{Line-of-Sight Dependent Quantities}

\begin{figure}
\vskip .4in
\plottwo{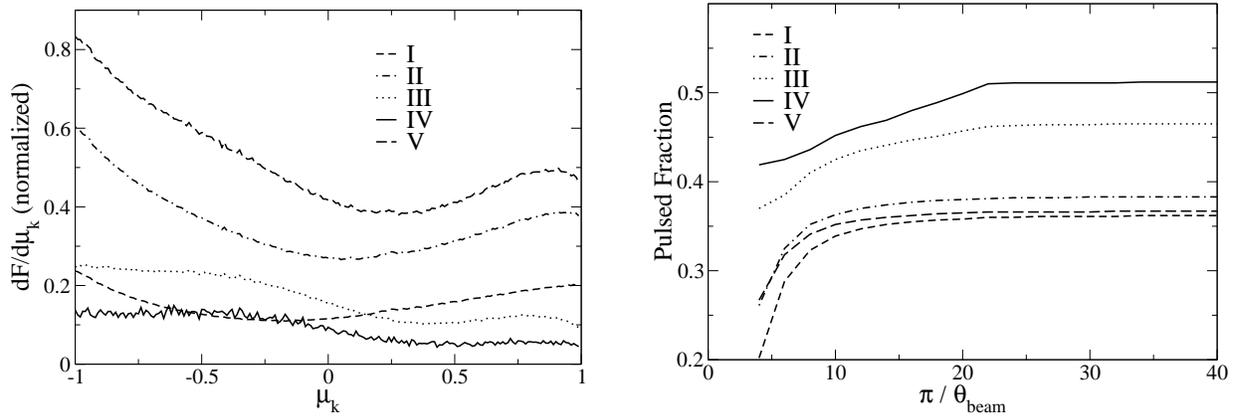}{f15b.eps}
\caption{\emph{Left:} Angular distribution of outgoing photons
(eq.~[\ref{eq:dFdmuk}]) in the 
five energy bands defined in eq.~(\ref{eq:energy_bands}). Bands III and IV contain most of the 
photons in the non-thermal tail of the spectrum, which is generated by multiple
resonant scattering. \emph{Right:} Pulsed fraction as function of beam angle $\theta_\mathrm{beam}$
obtained from the angular distributions shown in the left panel. Convergence to 1\% in all 
bands is achieved for $\theta_\mathrm{beam} = \pi/20 = 9^\circ$.}
\label{f:angdist}
\end{figure}

In this paper we calculate pulse profiles in different frequency 
bands (eq.~[\ref{eq:energy_bands}]), and along specific lines of sight.
Shown in the left panel of Fig.~\ref{f:angdist} is the angular distribution of
emission, eq.~(\ref{eq:dFdmuk}), for the same simulation shown 
in Figure~\ref{f:response}, assuming E-mode seed photons. 
These angular distributions are asymmetric with respect to $\mu_{\rm k} = 0$,
which allows for large pulsed fractions.  
The higher photon flux in the south magnetic hemisphere ($\mu_{\rm k} < 0$)
reflects the southward drift of the charges near the magnetic equator
(where the optical depth is largest).  The angular distribution in the
two lowest frequency bands also reflects the contribution of unscattered
seed photons, which peaks along the magnetic poles (see the right panel of 
Fig.~\ref{f:response}).  A much larger fraction of the photons in 
bands III and IV have undergone at least one scattering.

Because only a finite number of photons are evolved in each simulation,
a pulse profile can be obtained only by averaging the photon flux
over some angular cone of half-width $\theta_{\rm beam}$
(as in eq.~[\ref{eq:pulse_profile}]).   In reality, $\theta_{\rm beam}
\simeq R_{\rm res}/D \simeq 10^{-15}$ for a source of size
$R_{\rm res} \sim 100$ km (eq. [\ref{eq:rres}]) at a distance $D \sim 3$ kpc.  
In practice, $\theta_{\rm beam}$ is taken to be
the largest value for which the pulse profile has essentially converged
to a constant shape, and faithfully reproduces the result for 
$\theta_{\rm beam} \simeq 0$.  As a measure of convergence, we calculated
the pulsed fraction as a function of beam width for a few respresentative
profiles, and found convergence to within $\sim 1$ percent
at $\theta_\mathrm{beam} = \pi/20 = 9$~deg (see the right panel of 
Fig. \ref{f:angdist}).  In this calculation the angular grid is taken
to have a size $d\mu^\prime \simeq d\phi^\prime \simeq (1-\mu_\mathrm{beam})/2
\simeq (\theta_{\rm beam}/2)^2$, which yields a large number of angular cells
within the beam. 

\end{appendix}


\end{document}